\documentclass[10pt,a4paper]{article}
\usepackage[
  margin=2.5cm,
  includefoot,
  footskip=40pt,
]{geometry}

\usepackage{amsfonts}
\usepackage{url}

\usepackage{booktabs}   %% For formal tables:
\usepackage{subcaption} %% For complex figures with subfigures/subcaptions
\usepackage{xspace}     %% For spaces in macros,
\usepackage{extarrows}     %% arrows with text above

\usepackage{tikz}
\usetikzlibrary{quotes, shapes, positioning, fit, calc, decorations.pathreplacing}

\usepackage{listings}
\newtheorem{theorem}{Theorem}

\tikzset{
  register/.style={
    draw,
    text centered,
    anchor=mid,
    font=\ttfamily
  },
  archState/.style={
    outer sep=0,
  },
  uArchState/.style={
    draw,
    minimum width=15ex,
    minimum height=3.5ex,
    outer sep=0
  }
}

\def\reg#1#2{
  \node [register, label={[name=label]#1}](value) {#2};
  \node [inner sep=0, outer sep=0, fit=(label)(value)] (#1) {};
}

\newcommand{\archState}[8]{
  \begin{tikzpicture}
  \matrix(regs)[outer sep=0, node distance=0, ampersand replacement=\&, column sep=0mm] {
    \reg{r0}{#2} \& \reg{r1}{#3} #7 \& \reg{#6}{#4} #8 \& \reg{pc}{#5} \\
  };
  \end{tikzpicture}
}

\def\elipsis{
  \& \node(){...};
}

\def\reg#1#2{
  \node [register, label={[name=label]#1}](value) {#2};
  \node [inner sep=0, fit=(label)(value)] (#1) {};
}

\newsavebox\realArch
\newsavebox\microArch
\newsavebox\cache
\pgfdeclarelayer{background}
\pgfsetlayers{background,main}
\definecolor{lightGray}{RGB}{220,220,220}

\newcommand{\realArchState}[4]{
  \sbox{\realArch}{
    \archState{}{#1}{#2}{#3}{#4}{r2}{}{\elipsis}
  }
}

\newcommand{\uArchState}[4]{
  \sbox{\microArch}{
    \archState{}{#1}{#2}{#3}{#4}{r2}{}{\elipsis}
  }
}

\newcommand{\realArchStateSp}[4]{
  \sbox{\realArch}{
    \archState{}{#1}{#2}{#3}{#4}{sp}{\elipsis}{}
  }
}

\newcommand{\uArchStateSp}[4]{
  \sbox{\microArch}{
    \archState{}{#1}{#2}{#3}{#4}{sp}{\elipsis}{}
  }
}

\newcommand{\cacheState}[2]{
  \sbox{\cache}{
    \begin{tikzpicture}
    \matrix(state)[outer sep=0, node distance = 0, column sep=0mm, row sep=0mm, ampersand replacement=\&] {
      \node[outer sep=0]{P~}; \& \node[uArchState](){#1}; \\
      \node[outer sep=0]{C~}; \& \node[uArchState](){#2}; \\
    };
    \end{tikzpicture}
  }
}

\newcommand{\drawCpu}[0]{
  \begin{tikzpicture}[]
  \matrix(cpu)[column sep=0mm, row sep=0mm, ampersand replacement=\&] {
     \node(arch)[outer sep=0,archState]{\usebox{\realArch}}; \\
     \node(uArch)[outer sep=0,archState]{\usebox{\microArch}}; \\
     \node(cache)[outer sep=0]{\usebox{\cache}}; \\
  };   
  \begin{pgfonlayer}{background}
  \node[draw, fill=lightGray, inner sep=0, fit=(uArch)(cache)](uState){};
  \node[draw, inner sep=0, fit=(arch)(uState.north west)(uState.north east)](){};
  \end{pgfonlayer}
  \end{tikzpicture}
  \nobreak\hspace{-5ex}
}

\newcommand{\memLabel}[3]{
  \draw [decorate,decoration={brace,amplitude=2pt}]
         ($(#2.north east)+(2ex,-0.2ex)$) -- ($(#3.south east)+(2ex,0.2ex)$)
          node [anchor=west, midway, align=left, xshift=2pt]{#1};
}

\begin{document}

%% Title information
\title{Spectre is here to stay \\
  An analysis of side-channels and speculative execution}

%% Author information
%% Contents and number of authors suppressed with 'anonymous'.
%% Each author should be introduced by \author, followed by
%% \authornote (optional), \orcid (optional), \affiliation, and
%% \email.
%% An author may have multiple affiliations and/or emails; repeat the
%% appropriate command.
%% Many elements are not rendered, but should be provided for metadata
%% extraction tools.

\author{
  Ross Mcilroy\\
  Google\\
  \texttt{rcmilroy@google.com}          %% \email is recommended
  \and
  Jaroslav Sevcik\\
  Google\\
  \texttt{jarin@google.com}
  \and
  Tobias Tebbi\\
  Google\\
  \texttt{tebbi@google.com}
  \and
  Ben L. Titzer\\
  Google\\
  \texttt{titzer@google.com}
  \and
  Toon Verwaest\\
  Google\\
  \texttt{verwaest@google.com}
}

\date{\today}

\newcommand{\hsp}{\hspace{2pt}}
\newcommand{\hspm}{\hspace{5pt}}
\newcommand{\hspw}{\hspace{10pt}}
\newcommand\x[1]{\ensuremath{\mathit{#1}}\xspace}
\newcommand{\true}{\mathrm{T}}
\newcommand{\false}{\mathrm{F}}
\newcommand\K[1]{\ensuremath{\textsf{\sf#1}}\xspace}
\newcommand\KK[1]{\ensuremath{\K{\textbf{#1}}}\xspace}
\newcommand\Binop[1]{\ensuremath{\K{\textbf{#1}}}\xspace}
\newcommand\Binops[1]{\ensuremath{\K{\textbf{#1}}}s\xspace}

\maketitle

%% Abstract
%% Note: \begin{abstract}...\end{abstract} environment must come
%% before \maketitle command
\begin{abstract}
The recent discovery of the Spectre and Meltdown attacks represents a watershed moment not just for the field of Computer Security, but also of Programming Languages.
This paper explores speculative side-channel attacks and their implications for programming languages.
These attacks leak information through micro-architectural side-channels which we show are not mere bugs, but in fact lie at the foundation of optimization.
We identify three open problems, (1) finding side-channels, (2) understanding speculative vulnerabilities, and (3) mitigating them.
For (1) we introduce a mathematical meta-model that clarifies the source of side-channels in simulations and CPUs. 
For (2) we introduce an architectural model with speculative semantics to study recently-discovered vulnerabilities.
For (3) we explore and evaluate software mitigations and prove one correct for this model.
Our analysis is informed by extensive offensive research and defensive implementation work for V8, the production JavaScript virtual machine in Chrome.
Straightforward extensions to model real hardware suggest these vulnerabilities present formidable challenges for effective, efficient mitigation.
As a result of our work, we now believe that speculative vulnerabilities on today's hardware defeat all language-enforced confidentiality with no known comprehensive software mitigations, as we have discovered that untrusted code can construct a universal read gadget to read all memory in the same address space through side-channels.
In the face of this reality, we have shifted the security model of the Chrome web browser and V8 to process isolation.

\end{abstract}

%% 2012 ACM Computing Classification System (CSS) concepts
%% Generate at 'http://dl.acm.org/ccs/ccs.cfm'.
% <concept_id>10011007.10011006.10011008</concept_id>
% <concept_desc>Software and its engineering~General programming languages</concept_desc>
% <concept_id>10003456.10003457.10003521.10003525</concept_id>
% <concept_desc>Social and professional topics~History of programming languages</concept_desc>
% <concept_significance>300</concept_significance>

% \ccsdesc[500]{Software and its engineering~General programming languages}
% \ccsdesc[300]{Social and professional topics~History of programming languages}
%% End of generated code

%% Keywords
%% comma separated list
%\keywords{keyword1, keyword2, keyword3}  %% \keywords are mandatory in final camera-ready submission

%% \maketitle
%% Note: \maketitle command must come after title commands, author
%% commands, abstract environment, Computing Classification System
%% environment and commands, and keywords command.

\section{Introduction}

Computer systems aspire to enforce three key security properties: \emph{confidentiality}, \emph{integrity}, and \emph{availability}~\cite{BasicsOfInfoSec}.
Of these, confidentiality is the property that guarantees private information can only be accessed by authorized parties.
Confidentiality is the basis for many security mechanisms, including passwords, access tokens, TANs, cookies, and capabilities. 
It is enforced through a number of mechanisms, including encryption, spatial, temporal, or virtual separation, and mediation through access checks.
When ensuring confidentiality, it is not enough for implementations to be functionally correct.
In addition to computing the correct result, private data must not leak to unauthorized parties by any means.
Such leakage can happen through unforeseen information channels, called \emph{side-channels}.
Any measurable property of a computer implementation has the potential to be a side-channel.

Side-channels are an important area of study for computer systems and processes that deal with high-value secrets like crypto systems.
These attacks are not merely theoretical, as practical attacks have been demonstrated as far back as 1996~\cite{KocherAttack}. In high-security contexts, algorithms, software, and hardware are carefully designed to eliminate as many side-channels as possible.
This seems to be a never-ending battle, as attacks have been demonstrated using an amazing variety of measurements, including electromagnetic emissions ~\cite{agrawal2002side}, energy consumption~\cite{kocher1999differential}, power lines~\cite{PowerHammer}, microphones~\cite{AcousticSidechannel}, high resolution cameras~\cite{OpticalSidechannel}, IR photon detectors~\cite{PhotonicAttack}, and even the ambient light sensor of smartphones~\cite{spreitzer2014pin}.
Timing attacks use time itself as the measurable property, inferring confidential information by differential analysis of the varying time required to execute operations on or related to secret information. Timing was in fact one of the first side-channels used to attack crypto systems~\cite{KocherAttack}.

Timing side-channels are abundant because reducing execution time is an important goal of software and hardware engineering, and modern computer systems employ many optimizations which depend on the dynamic values of computations and thus might leak information about those dynamic values.
Confidentiality is at risk when code of different parties share resources such as a processor core, processor, processor package, memory bus, DRAM, cache, or disk.
Side-channels can exist both in time-sharing scenarios, where state persists between context switches, or in concurrent scenarios where processes directly contend for resources.

A classic example is processor caches which store the address and value of recently accessed memory locations.
Subsequent memory addresses are then faster for cached addresses, revealing information about which addresses were or were not recently accessed.
\cite{CacheAttackAES} was the first to use cache timing as a side-channel to attack crypto systems.
In principle, any processor unit with hidden state has the potential to store sensitive information and to leak this information to subsequently or concurrently executing code via a timing channel.
Outside of the CPU itself, DRAM can also be a side-channel~\cite{DramaUsenix}, since it contains active and inactive rows. 
\cite{ge2018survey} gives a more complete account of known attacks that use microarchitectural components. 

Security acknowledges that trust has levels.
Kernels are part of the trusted computing base, offering a platform for semi-trusted and trusted applications as isolated processes.
For the vast majority of installed applications, such as word processors or video games, we simply trust that they don't to steal our personal information and credentials, or attempt to impersonate us to third parties.
These attacks are however extremely serious for crypto systems and production settings when servers are virtualized on the same hardware, or the extreme case where multiple, competing, untrusted clients run on shared cloud computing platforms~\cite{CloudAttacks}.
In the case of a Web browser, where webpages run untrusted code and applications run semi-trusted code on devices close to us, trust is lower, and these attacks become relevant.
For the Web, untrusted JavaScript and WebAssembly code is sandboxed at multiple granularities to protect both the user's system and other webpages' private content.
As such, side-channels are now emerging as a real threat for the web platform.

\subsection{Security in Programming Languages}

Computer systems are massively complex, requiring oversimplification and reasoning via analogy. In this vein we can for the moment view them as a tower of abstractions that ultimate rests upon physics and builds upwards, from electrons to circuits, from circuits to micro-architectures, from micro-architectures to micro-ops, from micro-ops to ISAs, from ISAs to system software and libraries, and then programming languages, themselves an abstraction comprised of a some combination of a static compiler, runtime system, and virtual machine. When we consider with emulation and virtualization via hypervisors, the tower of hardware and software can include an unbounded number of emulation layers. Abstractions go further up from applications, of course, beyond a single computer to networked computer systems, clusters, databases, clouds, the internet, the Web.

In this view, programming languages, as most people use them, are square in the middle of this tower of abstractions, though they are relevant to upper layers that are programmable. Being the abstraction in which algorithms are expressed means the security properties of a programming language are therefore paramount to the security of algorithms and systems implemented with it. Security at this layer is important, since many decades of experience have shown us that secure hardware is not enough and securing software has many remaining challenges.

Although it is rarely explicitly recognized by name, most programming languages enforce various degrees of both integrity and confidentiality for programmer-created abstractions, preventing bugs and unpredictable results and increasing our confidence in software correctness. Basic programming language mechanisms such as proper scoping, namespace separation, storage separation, dynamic bounds checks, and both static and dynamic type checks are all attempts to prevent one part of a program or library from accessing another's data inappropriately. Of course, some languages are more explicitly security-conscious than others. For example some languages express confidentiality explicitly~\cite{sabelfeld2003language}. An early example is Jif~\cite{myers2001jif} which extends Java with a type system that expresses principals and access rights and extends the Java type system to ensure that data cannot be read by principals that were not granted access rights. 

In mainstream modern languages, strong type systems are designed to guarantee that a program cannot exhibit certain dangerous behaviors which might jeopardize integrity, confidentiality, and availability.
A strong type system allows reasoning about programs manipulating sensitive data.
Memory safety is among the crucial properties a strong type system must enforce.
In principle, memory-safe languages can guarantee strong isolation similar to process isolation as offered by common hardware and operating systems.
There even have been serious efforts~\cite{hunt2007singularity,nightingale2009helios} to build operating systems where processes share the same address space and are only isolated by a memory-safe programming language.
In this case, the burden of ensuring confidentiality rests fully on the type system, compiler, and runtime system rather than on hardware/OS process isolation.
A language implementation may enforce security properties through a combination of static and dynamic checks.

Language security is a powerful tool, but what if a programming language is run on insecure hardware? What if the CPU, even if verified to be architecturally correct~\cite{Sawada2002}, allows the program to bypass language security measures through side channels, which are outside architectural models? Our proofs of type safety and of type system soundness, which help us enforce confidentiality through integrity of abstractions, taint freedom, and other things, would suddenly be called into question.
In short, that would be bad.

\subsection{A new and pervasive threat}

To date, timing attacks and side-channels have only been used to observe \emph{legitimate} computations that happened in the ordinary course of a program's execution.
Thus the risk for information leakage could be determined by considering the algorithm and the data it processes. In turn, this was mostly seen as a risk for encryption algorithms and other programs doing heavy computations on sensitive data. This offered the possibility to selectively harden sensitive algorithms.

Information leaks from speculative execution represent a different level of threat that we must now begin to understand and reason about.
Spectre~\cite{GoogleProjectZero,Kocher2018spectre} allows for information to be leaked from computations that should have never happened according to the architectural semantics of the processor.
Spectre attacks use targeted manipulations of the shared microarchitectural state to trigger ordinarily impossible computations, either in untrusted code on which an implementation imposes safety checks, or by injecting dangerous speculative behavior into trusted code.
In either case, Spectre allows for low-level read access to all of the addressable memory. This puts arbitrary in-memory data at risk, even data ``at rest" that is not currently involved in computations and was previously considered safe from side-channel attacks.

This paper is an attempt to distil and clarify that threat. As a result of our\footnote{See acknowledgments section} work on Spectre, we now know that information leaks may affect all processors that perform speculation, regardless of instruction set architecture, manufacturer, clock speed, virtualization, or timer resolution. Since the initial disclosure of three classes of speculative vulnerabilities, all major vendors have reported affected products, including Intel, ARM, AMD, MIPS, IBM, and Oracle. This class of flaws are deeper---at the microarchitectural level of the processor---and more widely distributed---in essentially every high performance processor---than perhaps any security flaw in history, affecting billions of CPUs in production across all device classes. While speculation is often informally equated with branch prediction, the concept of speculation is broader, since processors speculate in other ways not related to branch prediction, as we show in section \ref{sec:variant4}. Vulnerabilities from speculative execution are not processor bugs but are more properly considered fundamental design flaws, since they do not arise from errata. Troublingly, these fundamental design flaws were overlooked by top minds for decades. Our paper shows these leaks are not only design flaws, but are in fact foundational, at the very base of theoretical computation. On the opposite end of the spectrum, we detail our practical attempts at software mitigation in a production virtual machine entrusted with the security of the Web.

\vspace{10ex}
\subsection{Contributions}

\begin{itemize}
\item A formal model of microarchitectural side-channels and optimizations
\item An abstract amplification technique to make even the smallest timing difference detectable with any resolution clock
\item A formal model that explains several recently-discovered classes of speculative vulnerabilities
\item A succinct description of the ultimate leak from speculation: the universal read gadget
\item Construction of the universal read gadget using multiple vulnerabilities
\item An analysis of exploitable source language features
\item A formal analysis of mitigations for the variant 1 vulnerability
\item Discussion of mitigations for other variants
\end{itemize}

\section{Understanding microarchitectural side-channels} \label{sec:u-state}

\newcommand\Ints{\ensuremath{\mathbb{Z}}\xspace}
\newcommand\Nats{\ensuremath{\mathbb{N}}\xspace}
\newcommand\Reals{\ensuremath{\mathbb{R}}\xspace}
\newcommand\Bools{\ensuremath{\mathbb{B}}\xspace}

\newcommand{\Arch}[2]{{\Lambda \langle #1, #2\rangle}}
\newcommand{\MuArch}{$\mu$-architecture\xspace}
\newcommand{\MuArchs}{$\mu$-architectures\xspace}
\newcommand{\MuArchal}{$\mu$-architectural\xspace}
\newcommand{\MuArchly}{$\mu$-architecturally\xspace}
\newcommand{\MuState}{$\mu$-state\xspace}
\newcommand{\MuStates}{$\mu$-states\xspace}
\newcommand{\MuStep}{$\mu$-step\xspace}
\newcommand{\MuSteps}{$\mu$-steps\xspace}
\newcommand{\MuStateType}{\ensuremath{\Delta}\xspace}
\newcommand{\MuExtension}{$\mu$-extension\xspace}
\newcommand{\True}{\text{\texttt{true}}\xspace}
\newcommand{\False}{\text{\texttt{false}}\xspace}
\newcommand{\One}{\text{\texttt{1}}\xspace}
\newcommand{\Zero}{\text{\texttt{0}}\xspace}

\newcommand\Bits{\ensuremath{\{\Zero, \One\}}\xspace}

It is often convenient to think of a CPU as an interpreter which executes instructions, maintaining a program counter, registers, and memory along the way, a collection of information which together comprise what is known as the \emph{architectural} state of a program.
Yet in reality, all CPUs are simulators; they are implemented as complex circuits organized into register files, functional units, tables, forwarders, caches, and many other things, each with their own internal administrative state, and only simulate the action of an interpreter.
A CPU's internal state, often called the \emph{microarchitectural} state, abbreviated here \emph{\MuState}, is hidden by the abstraction that the CPU provides to programs.
However, as we'll see in this section, this abstraction is more permeable than it would at first appear, once we consider the ability to measure execution time.

For the moment, let's step back from concrete computers and introduce a mathematical meta-model to make reasoning about state-transition systems both abstract and rigorous.
The meta-model allows us to reason about any kind of computational system, not just CPUs, Turing machines, or equivalent calculi.
We use a simulation relation that relates architectural states to \MuArchal states, allowing many correct \MuStates to map to a single architectural state.
It is exactly in this differential mapping where side-channels can occur.
This model is deterministic and requires only a single event counter, or timer, to extract information from the side-channel.
This shows that these vulnerabilities are \emph{not} the result of complex nondeterministic machines or specific CPU bugs, but are in fact, fundamental to simulation in general.

\hfill

\newcommand\archz[1]{\ensuremath{\mathit{\overline{#1}}}\xspace}
\newcommand\archp[2]{\ensuremath{\mathit{\overline{#1}_{#2}}}\xspace}

\textbf{Architecture} We define an \emph{architecture} \archz{\alpha} $: P \times \Sigma \rightarrow \Sigma$ as a computable function
$\archp{\alpha}{\rho}(\sigma) = \sigma'$
 where $\rho \in P$ represents a program, and $\sigma \in \Sigma$ represents the state of the program. We denote an architecture for programs $P$ with states $\Sigma$ by $\Arch{P}{\Sigma}$. We make no restrictions on the language of programs $P$ or the language of states $\Sigma$ as long as they are countable sets. In the next section it will be convenient to mimic a real CPU by modeling memory cells, but for now we allow P and $\Sigma$ to encode any kind of computation, e.g. with syntactic terms like $\lambda x . x$ and reduction rules. Further, we assume nothing about how $\archz{\alpha}$ is most succinctly described, whether it be mathematically with a small-step semantics, operationally with a state transition diagram, or procedurally as a set of instructions for yet another architecture.

\hfill

\textbf{$\mu$-Architecture} We define the execution semantics of a \emph{\MuArch} for $\archz{\alpha}: \Arch{P}{\Sigma}$ as an architecture $\archz{m}: \Arch{P}{\MuStateType}$ where we choose two sets of \emph{observable} states $\Omega_\Sigma \subseteq \Sigma$ and $\Omega_\MuStateType \subseteq \MuStateType$ such that there exists a simulation relation function $R: \Omega_\MuStateType \rightarrow \Omega_\Sigma$ with:

\begin{itemize}
\item $\forall \sigma \in \Omega_\Sigma$, $\exists \mu \in \Omega_\MuStateType$ such that $R(\mu) = \sigma$ 
\newline
meaning that all observable architectural states have at least one corresponding observable \MuArchal state (i.e. the relation is \emph{surjective}).
\end{itemize}

The relation $R$ may map many $\mu_i$ to a single $\sigma$, allowing $\archz{m}$ the freedom to add additional details like a memory cache or branch predictor state. It is exactly this extra state that can exhibit side-channels.

We can now precisely define \emph{correctness} for an $\archz{\alpha}$ and its implementation $\archz{m}$ as a condition on execution paths between \emph{observable related} states. 

\begin{itemize}
\item $\forall n_1 \in \Nats, \sigma, \sigma' \in \Omega_\Sigma$ such that $\archp{\alpha}{\rho}^{n_1}(\sigma) = \sigma'$, $\exists n_2 \in \Nats, \mu, \mu' \in \Omega_\MuStateType$ such that $R(\mu) = \sigma$, $\archp{m}{\rho}^{n_2}(\mu) = \mu'$, and $R(\mu') = \sigma'$
\newline meaning that for any architectural execution path between two observable states, there is a \MuArchal execution path between correspondingly related observable \MuStates.
\end{itemize}

Since architectures are deterministic state transition systems, we can make use of the existential quantifier for the correctness property\footnote{
In essence, for each observable starting state, we only require there is \emph{some} \MuArchal state from which computation can begin, and determinism takes over. Universally quantifying the path length accomplishes induction for full correctness. This choice allows \MuArchs that can even skip observable states under some circumstances, as long as they can be configured to reach \emph{any desired} observable state reachable by the architectural execution.}.

\subsection{Writing information into \MuState}

The additional state of a \MuArch might be used by a crafty program to encode information which is \emph{hidden} from the architectural state. To see this, consider an architecture $\archz{\alpha}: \Arch{P}{\Sigma}$ and implementation $\archz{m}: \Arch{P}{\MuStateType}$. Consider a program $w \in P$ that receives as \emph{input} either \Zero or \One and writes this bit into the \MuArchal state and then ``forgets" it. More precisely:

\begin{itemize}
\item $\forall b \in \Bits$ let $\Sigma_{w}(b) \subset \Omega_\Sigma$ be the architectural states encoding $b$ as input to $w$ 
\newline with $\Sigma_{w}(\Zero) \cap \Sigma_{w}(\One) = \emptyset$
\item $\exists \phi \in \Omega_\Sigma, \forall \sigma \in \Sigma_{w}(b), \exists n$ such that $ \archp{\alpha}{w}^{n}(\sigma) = \phi$
\newline meaning that all inputs representing \Zero or \One converge to the same observable state $\phi$
\item $\exists V \in \Omega_\MuStateType \rightarrow \Bits$ such that 
\begin{itemize}
\item $\forall b \in \Bits, \sigma \in \Sigma_{w}(b), \mu \in \Omega_\MuStateType $ such that $ (\sigma, \mu) \in R, $ let $k \in \Nats$ be the smallest number such that  $\archp{m}{w}^{k}(\mu) = \mu_\phi$ and $(\phi, \mu_\phi) \in R,$ then $V(\mu_\phi) = b$
\newline meaning that the shortest\footnote{Restricting to shortest paths thus only requires the bit be viewable for a non-zero time window immediately after having reached $\phi$ the first time. This handles the general case of infinite loops involving $\phi$ and machines where the \MuArchly-encoded bit can be flipped or lost.} \MuArchal paths that lead observable states related to $\phi$ preserve the input bit $b$ and can be inspected by the function $V$.
\end{itemize}
\end{itemize}

We encode the input bit $b$ by starting with a $\sigma$ chosen from the appropriate set.
By design, when $w$ executes, it converges on a common state $\phi$, thus ``forgetting" $b$ at the architectural level.
However, $w$ organizes the output \MuStates into two disjoint sets, and with the function $V$, $b$ can be recovered.
Thus the program $w$ constitutes a \emph{write mechanism} for a \MuArchal side-channel.

\subsection{Reading information from \MuState} \label{sec:write-mechanism}

A correct implementation of an architecture is a complete abstraction: even with a write mechanism to store information in \MuArchal state, the information is, by design, inaccessible at the architectural level.
However, many architectures offer ways to measure some aspect of the \MuState.
This can be a direct measurement (e.g. is a given memory address in cache) or some approximation of the execution history (e.g. a \emph{timer} or an \emph{event counter}).
These mechanisms may violate the architectural abstraction and allow a crafty program to create a \emph{read mechanism} to complete the construction of a side-channel.

\textbf{Events.}
For an architecture $\archz{\alpha}: \Arch{P}{\Sigma}$, we define an \emph{event} as a predicate function $P_E : P \times \Sigma \rightarrow \Bits$ that determines if a special situation $E$ has occurred at a given state. It is straightforward for a program to count its own events, maintaining a counter  encoded in $\Sigma$. Such architectural event counters pose no threat to the integrity of the architectural abstraction. However, very often a \MuArch counts \MuArchal events (i.e. with a predicate over \MuStateType) and offers an API function $f_E : () \rightarrow \Nats$ to return the counter value. This does indeed break the abstraction, allowing programs to construct the read mechanism to complete the side-channel.

\textbf{Clocks.}
It's easy to see that clocks are simply special cases of event counters. The basic building block of all clocks is the \emph{step counter} (i.e. $P_E(\rho, \mu) = \One$) which advances for every step of the \MuArch execution. Of course a clock may advance at a different rate than the step counter and potentially drift. To model this, we define a \emph{clock} $C: \Nats \rightarrow \Nats$ as a linear, monotonically increasing function from the step counter:
\begin{itemize}
\item $C(0) = 0$\newline
All clocks start at $0$ initially.
\item $C(n+1) \geq C(n)$\newline
Each reading is greater than or equal to the last (monotonically increasing).
\item $\exists r \in \mathbb{R} > 0$ and $B \in \Nats$ such that $\forall n  \hspm |C(n) - rn| \leq B$\newline
Every clock has a nonzero average rate $r$ and a drift bound $B$ such that every reading is within $B$ of the linear function $rn$.
\end{itemize}

Thus a clock measures the passage of time as an ever-increasing number, never runs backwards, and always runs at more or less the same rate, no matter how long we use it.

\subsection{A timing channel that exploits optimizations} \label{sec:opts}

CPUs and simulators take shortcuts to improve the execution time of programs, making use of mechanisms such as caches and branch predictors.
A common theme among most such optimizations is that they depend only on \MuState: a single architectural state can result in different execution time, depending on whether the \MuState is ``fast" or ``slow".
This differential is the basis of a \emph{timing channel}, as a program intentionally manipulates \MuState to trigger either fast or slow executions.

A timing channel, like all information channels, is comprised of separate \emph{read} and \emph{write} mechanisms.
It is easy to construct a program whose execution time varies by any given amount by simply having the program do completely different things depending on the input.
However, we are interested in timing channels at the \MuArchal level, where programs that exhibit no visible architectural divergence depending on their input, nevertheless exploit \MuArchal differences for timing channel construction.
Such timing channels are not visible without a \MuArchal model.

These can be described as follows:

\newcommand\Opt{\texttt{opt}\xspace}
\newcommand\Trig{\texttt{trig}\xspace}

\begin{itemize}
\item \textbf{Write mechanism}. We will use the term \emph{optimization trigger} to denote a program fragment \Trig that intentionally introduces a difference in the \MuState that will lead to different execution time for \emph{another program}, \emph{in the future}. As in \ref{sec:write-mechanism}, \Trig takes as input a bit $b$ and organizes \MuStates such that a function $V$ can recover the bit. 
\item \textbf{Read mechanism}. We will use the term \emph{optimization opportunity} to denote a program fragment \Opt whose execution time varies depending on its input \MuState. By executing \Opt, information is transferred from the \MuState domain into the \emph{timing domain}. To complete the mechanism, i.e. to finish the implementation of $V$, we read the information from the timing domain back into the \emph{architectural domain} by simply reading the clock and deciding if execution was fast or slow.
\end{itemize}

\subsection{Amplification Lemma} \label{sec:amp-lemma}

As we have now seen, the last step of a timing channel's read mechanism is to read the clock.
However, a clock may in fact have lower resolution than variation in \MuSteps due to optimizations.
For example, an L1 cache hit on today's fastest CPUs may take 3 cycles, while a miss to L2 may take 12 or more cycles.
For a processor running at 3Ghz, this difference of 9 cycles is therefore 3 nanoseconds, much smaller than the 1000-nanosecond-resolution clocks typically offered by most programming language APIs.
Is the information therefore gone, inaccessible to programs?

The answer is, unfortunately, no.
The information remains as \emph{fractional bits} stored in the timing domain.
These fractional bits can be made visible with a lower resolution clock using a variety of amplification techniques, e.g. by simply batching multiple optimization triggers, or repeating the trigger+optimization combination as many times as needed.

\textbf{Lemma.} Suppose we have an architecture \archz{a} and implementation \archz{m}.

\begin{itemize}
\item let \Trig represent an optimization trigger and \Opt represent an optimization opportunity.
\item let $r$ and $B$ be the clock rate and drift bound of the clock available in \archz{m}.
\item let $\delta$ be the timing difference, in \MuSteps, between the fastest execution of \Opt for input \Zero and the slowest execution for input \One.
\item let $rB = \x{ceiling}(\x{max}(B, B/r) / \delta)$.
\item let $\rho_{rB} = Compile($\texttt{"}.
{\small
\begin{lstlisting}[mathescape]
      if (input == 0) {
         for (i = 0; i < rB; i++) {
           $\Trig(\Sigma_{\Trig}(0)$)
           $ \Opt()$
         }
      } else {
         for (i = 0; i < rB; i++) {
           $\Trig(\Sigma_{\Trig}(1))$
           $ \Opt()$
         }
      }
      if (timer() > threshold) return 1;
      return 0;\end{lstlisting}}
\hsp \texttt{"}$)$
\end{itemize}

The above program executes the same number of architectural steps, no matter its input.
However, it amplifies the \MuArchal timing differences due to \Trig and \Opt to be detectable with any clock, if we know the clock rate and its drift bound.
Since the timing domain is cumulative, we can repeatedly write fractional bits into the timing domain, and then when we are sure their sum is more than the clock resolution, we can extract a whole bit with a single clock reading.

\textbf{Generality.} We argue that nearly all optimizations that could be performed by a \MuArch are observable using this amplification technique or a similar one.
The argument ultimately rests on the ability to repeat an optimization arbitrarily many times.
Intuitively this is always possible since \emph{by design}, optimizations are intended to improve the asymptotic runtime of a program (i.e. make it $N$\% faster), no matter how long it runs.
An optimization that shuts off after some number of repetitions achieves only a constant speedup (i.e. a maximum of $K$ seconds faster), which approaches $0$\% the longer a program runs.
We can also argue directly from the technical details of optimizations.
Since optimizations are based on \emph{local} \MuState observations, including local history, the extra state needed to maintain a loop can always be separated from the state that triggers the optimization, forcing the optimization to repeatedly occur.

\textbf{Impossibility of complete mitigation with timers.} Based on the generality argument, we argue that mitigating timing channels by manipulating timers is impossible, nonsensical, and in any case ultimately self-defeating.
For example, a common thought is that perhaps the \MuArch can track all time that has been saved due to optimizations and somehow charge the program back.
To see why this does not work, first, we require the timer to track the \MuArchal steps as a clock with bounded drift, so the \MuArch cannot lie forever about the clock.
Instead, it must regularly charge the program back, e.g. by \emph{waiting} in order to stay with the drift bound. Such a scheme is nonsensical because it simply wastes \MuSteps\footnote{Also note that waiting might consume less power in real hardware, risking another side-channel.}. We argue that elaborate charge-back systems like this are equivalent to not doing optimizations at all; they approach a constant time implementation with resolution of the drift bound. As such, the asymptotic performance benefit from optimizations is again $0$. Indeed, it may be case that constant-time implementations may be the \emph{only way} to avoid leaks via timing side-channels. Timer coarsening as a mitigation at best lowers the effective bandwidth of a timing channel and simply increases the complexity of the read and write mechanisms without making them impossible.

\section{An Architecture to study Spectre}

In the last section, we proved that any \MuArchal optimization is ultimately observable at the architectural level through timing, showing a general amplification technique to construct programs that repeatedly trigger the same optimization in order to amplify its effect. In that exercise we used a meta-model to show that information leaks affect all models of computation. In this section we introduce a series of semantic models to study the problem of information leaks due to speculative execution like that in today's CPUs. Note that we don't model multi-core systems, threads, pipelines, or memory barriers, as these are not necessary to demonstrate speculative vulnerabilities.

In Figure~\ref{fig:arch_model}, we introduce the language of programs and states and the execution semantics of programs. We will use this architecture for the remainder of this section.

%TODO: add a figure and give our architecture a name

\newcommand\StmtF[1]{\ensuremath{\textsf{\sf#1}}}
\newcommand{\StmtNop}{\StmtF{nop}}
\newcommand{\StmtConst}{\StmtF{const}}
\newcommand{\StmtLoad}{\StmtF{load}}
\newcommand{\StmtStore}{\StmtF{store}}
\newcommand{\StmtOp}{\StmtF{op}}
\newcommand{\StmtBinop}{\StmtF{binop}}
\newcommand{\StmtLabel}{\StmtF{label}}
\newcommand{\StmtBranch}{\StmtF{branch}}
\newcommand{\StmtJump}{\StmtF{jump}}
\newcommand{\StmtCall}{\StmtF{call}}
\newcommand{\StmtReturn}{\StmtF{return}}

\newcommand{\srule}[3]{\dfrac{#1}{#2}\ \ \textrm{#3}}
\newcommand{\nj}[3]{#1, #2 \longrightarrow #3}
\newcommand{\sj}[3]{#1, #2 \xrightarrow{\mathrm{spec}} #3}
\newcommand{\sjstar}[3]{#1, #2 \xrightarrow{\mathrm{spec}}_* #3}
\newcommand{\supdate}[3]{#1[#2] \leftarrow #3}

\newcommand{\concat}{+\!\!\!\!\!+}
\newcommand{\predict}{\mathrm{predict}}
\newcommand{\invert}{\mathrm{invert}}
\newcommand{\naive}{na\"{i}ve\xspace}
\newcommand{\Naive}{Na\"{i}ve\xspace}
\newcommand{\updatePrediction}{\mathrm{updatePrediction}}

\newcommand\IConst{\KK{const}}
\newcommand\ILoad{\KK{load}}
\newcommand\IStore{\KK{store}}
\newcommand\IBinop{\KK{binop}}
\newcommand\IBranch{\KK{branch}}
\newcommand\IJump{\KK{jump}}
\newcommand\ITimer{\KK{timer}}

\newcommand\Imm{\ensuremath{\#\Ints}}

\newcommand{\lookupR}{\mathrm{lookupR}}
\newcommand{\lookupM}{\mathrm{lookupM}}
\newcommand{\nospec}{\mathrm{nospec}}

\begin{figure}
\centering
\begin{minipage}{.3\textwidth}
{\small
\vspace{2ex}
$$
\begin{array}{@{}r@{~}c@{~}l@{}}
\multicolumn{3}{l}{\text{(registers)}} \\
 \x{r} &::=&
  \K{r0} ~|~
  \ldots ~|~
  \K{r15} \\
\multicolumn{3}{l}{\text{(operations)}}\\
  \x{op} &::=&
  \Binop{eq} ~|~
  \Binop{ge} ~|~ \\&&
  \Binop{add} ~|~
  \Binop{sub} ~|~ \\&&
  \Binop{mul} ~|~ 
  \Binop{div} ~|~ \\&&
  \Binop{shl} ~|~
  \Binop{shr} ~|~ \\&&
  \Binop{and} ~|~ 
  \Binop{or} ~|~ \\&&
  \Binop{xor} \\
\multicolumn{3}{l}{\text{(instructions)}} \\
  \x{i} &::=&
  \IConst \hsp r_d \hsp \Imm ~|~ \\&&
  \ILoad \hsp r_d \hsp [r_a + \Imm]  \hsp ~|~ \\&&
  \IStore \hsp [r_a + \Imm] \hsp r_v  \hsp ~|~ \\&&
  \IBinop \hsp \mathit{op} \hsp r_d \hsp r_a \hsp r_b \hsp ~|~\\&&
  \IBranch \hsp r_c \hsp \#\Ints  \hsp ~|~\\&&
  \IJump \hsp r_d \\&&
  \ITimer \hsp r_d \\
\multicolumn{3}{l}{\text{(program $\equiv$ P)}} \\
  \x{\rho} &::=& \vec{i} \\
\multicolumn{3}{l}{\text{(states $\equiv \Sigma$)}} \\
  \hspace{10ex}\x{\sigma} &::=& \langle\x{pc}: \Ints, \hsp \x{R}: \vec{\Ints}, \hsp \x{M}: \vec{\Ints}\rangle
\end{array}
$$
}

\end{minipage}
~
\begin{minipage}{.65\textwidth}
{\small
  \begin{align*}
    \srule{\rho[\x{pc}] = \IConst \hsp r_d \hsp k}
          {\nj{\rho}{\langle \x{pc}, R, M \rangle}
                 {\langle \x{pc}+1, \supdate{R}{r_d}{k}, M \rangle}}
          {Arch-Const}
    \\[1em]
    \srule{\rho[\x{pc}] = \ILoad \hsp r_d \hsp [r_a + \x{k}]}
          {\nj{\rho}{\langle \x{pc}, R, M \rangle}
                 {\langle \x{pc}+1, \supdate{R}{r_d}{M[R[r_a] + \x{k}]}, M \rangle}}
          {Arch-Load}
    \\[1em]
    \srule{\rho[\x{pc}] = \IStore \hsp [r_a + \x{k}] \hsp r_v}
          {\nj{\rho}{\langle \x{pc}, R, M \rangle}
                 {\langle \x{pc} + 1, R, \supdate{M}{R[r_a] + \x{k}}{R[r_v]} \rangle}}
          {Arch-Store}
    \\[1em]
    \srule{\rho[\x{pc}] = \IBinop \hsp \x{op} \hsp r_d \hsp r_a \hsp r_b}
          {\nj{\rho}{\langle \x{pc}, R, M \rangle}
                 {\langle \x{pc} + 1, \supdate{R}{r_d}{\x{op}(R[r_a], R[r_b])}, M \rangle}}
          {Arch-Binop}
    \\[1em]
    \srule{\rho[\x{pc}] = \IBranch \hsp r_c \hsp \x{d} \hspw R[r_c] \neq 0}
          {\nj{\rho}{\langle \x{pc}, R, M \rangle}
                 {\langle \x{pc}+\x{d}, R, M \rangle}}
          {Arch-BranchTaken}
    \\[1em]
    \srule{\rho[\x{pc}] = \IBranch \hsp r_c \hsp \x{d} \hspw R[r_c] = 0}
          {\nj{\rho}{\langle \x{pc}, R, M \rangle}
                 {\langle \x{pc}+1, R, M \rangle}}
          {Arch-BranchNotTaken}
    \\[1em]
    \srule{\rho[\x{pc}] = \IJump \hsp r_d}
          {\nj{\rho}{\langle \x{pc}, R, M \rangle}
                 {\langle R[r_d], R, M \rangle}}
          {Arch-Jump}
    \end{align*}
}
\end{minipage}
\caption{Architectural model}
\label{fig:arch_model}
\end{figure}

\begin{itemize}
\item A program $\rho$ consists of a vector $\vec{i}$ of instructions indexed with an integer program counter.
\item The architectural states $\sigma$ consist of a triple ${\langle \x{pc}, R, M \rangle}$ of program counter $\x{pc} \in \Ints$, a vector $\x{R} \in \vec{\Ints}$ of registers, and vector $\x{M} \in \vec{\Ints}$ representing memory.
\item A \IBranch has an input condition register $r_c$ and a relative pc offset.
\item A \IJump is an unconditional indirect jump to the pc contained in the register $r_d$.
\end{itemize}

\subsection{Microarchitecture}

Our architecture is simple yet complete enough to illustrate several broad of classes vulnerabilities that arise from speculative execution. We will now begin enhancing this architecture with models of \MuArchal state in order to study how speculative vulnerabilities arise.

\subsubsection{Modeling memory caches}

\newcommand\Max[1]{{\ensuremath{#1_\x{max}}}}

We extend the architectural state of the program $\Sigma$ by adding \MuState $C: \Ints^\Max{C}$ a list of cached memory addresses of maximum fixed capacity $\Max{C} > 0$ which holds the addresses currently stored in the cache. We add a cache-update function $\x{LRU}_n: (\Ints^n, \Ints) \rightarrow \Ints^n$ which models adding a new memory address to the cache and evicting the least-recently-used\footnote{The exact cache replacement algorithm is mostly orthogonal to this work. We use LRU here because it is simple to model.} address if necessary to limit the entries to $n$. Evaluation rules are as before, with the following modifications to the rules for \ILoad and \IStore:

%TODO(titzer): add rules for LRU_n

{\small
  \begin{align*}
    \srule{\rho[\x{pc}] = \ILoad \hsp r_d \hsp [r_a + \x{k}] \hspw R[r_a] + \x{k} \notin C \hspw C' = \x{LRU}_\Max{C}(C, R[r_a] + \x{k})}
          {\nj{\rho}{\langle \x{pc}, R, M, C \rangle}
                 {\langle \x{pc}, R, M, C' \rangle}}
          {Uncached-Load}
    \\[1em]
    \srule{\rho[\x{pc}] = \IStore \hsp [r_a + \x{k}] \hsp r_v \hspw R[r_a] + \x{k} \notin C \hspw C' = \x{LRU}_\Max{C}(C, R[r_a] + \x{k})}
          {\nj{\rho}{\langle \x{pc}, R, M, C \rangle}
                 {\langle \x{pc}, R, M, C' \rangle}}
          {Uncached-Store}
    \\[1em]
    \srule{\rho[\x{pc}] = \ILoad \hsp r_d \hsp [r_a + \x{k}] \hspw R[r_a] + \x{k} \in C \hspw C' = \x{LRU}_\Max{C}(C, R[r_a] + \x{k})}
          {\nj{\rho}{\langle \x{pc}, R, M, C \rangle}
                 {\langle \x{pc}+1, \supdate{R}{r_d}{M[R[r_a]]}, M, C' \rangle}}
          {Cached-Load}
    \\[1em]
    \srule{\rho[\x{pc}] = \IStore \hsp [r_a + \x{k}] \hsp r_v \hspw R[r_a] + \x{k} \in C \hspw C' = \x{LRU}_\Max{C}(C, R[r_a] + \x{k})}
          {\nj{\rho}{\langle \x{pc}, R, M, C \rangle}
                 {\langle \x{pc} + 1, R, \supdate{M}{R[r_a]}{R[r_v]}, C' \rangle}}
          {Cached-Store}
    \end{align*}
}

That is, we require that loads and stores first load their addresses into the cache, penalizing uncached accesses with an extra \MuStep.

\subsubsection{Branch Predictor State}

Branch prediction is a far more complex process than adding a cache, since the CPU begins to execute instructions \emph{speculatively} before a branch outcome is known, and discard the architectural effects if the speculation is wrong. However, the \MuState to feed branch prediction is relatively simple. We add \MuState $B: (\Ints, \Bools)^\Max{B}$ a list of program counter location/prediction pairs of maximum fixed capacity $\Max{B} > 0$. We add a prediction-update function $\x{BP_n} : ((\Ints, \Bools)^n, \Bools) \rightarrow (\Ints, \Bools)^n$ that updates the prediction for a branch given its outcome. The prediction state is updated after a branch outcome is known. It will be used later in section \ref{sec:modeling-control-speculation}. 

{\small
  \begin{align*}
    \srule{\rho[\x{pc}] = \IBranch \hsp r_c \hsp \x{d} \hspw R[r_c] \neq 0 \hspw B' = \x{BP_{\Max{B}}}(B, \True)}
          {\nj{\rho}{\langle \x{pc}, R, M, B \rangle}
                 {\langle \x{pc}+\x{d}, R, M, B' \rangle}}
          {Record-BranchTaken}
    \\[1em]
    \srule{\rho[\x{pc}] = \IBranch \hsp r_c \hsp \x{d} \hspw R[r_c] = 0 \hspw B' = \x{BP_{\Max{B}}}(B, \False)}
          {\nj{\rho}{\langle \x{pc}, R, M, B \rangle}
                 {\langle \x{pc}+1, R, M, B' \rangle}}
          {Record-BranchNotTaken}
    \end{align*}
}

\subsubsection{Indirect Branch Predictor State}

We model indirect branch prediction by additional \MuState $J: (\Ints, \Ints)^\Max{J}$ a list of program counter location pairs of maximum fixed capacity $\Max{J} > 0$. While similar to the branch predictor state, this jump table, which is often referred to as the \emph{branch target buffer} in computer architecture literature, stores a target address for each indirect jump. As with the branch prediction state, we add a function $JP_n : ((\Ints, \Ints)^n, \Ints) \rightarrow (\Ints, \Ints)^n$ which updates the jump table with the target address for a given jump after it is executed.

{\small
  \begin{align*}
    \srule{\rho[\x{pc}] = \IJump \hsp r_d \hspw J' = JP_{\Max{J}}(J, R[r_d])}
          {\nj{\rho}{\langle \x{pc}, R, M, J \rangle}
                 {\langle R[r_d], R, M, J' \rangle}}
          {Record-Jump}
    \end{align*}
}

\subsubsection{Modeling Control Speculation}\label{sec:modeling-control-speculation}

To model control speculation, we further augment the state with a reorder buffer which is a sequence of instructions that are waiting to be evaluated and will later be committed to the architectural state. The reorder buffer models two aspects of speculative execution:
\begin{itemize}
  \item Instructions are evaluated out-of-order, after their input dependencies are executed.
  \item Branch outcome can be predicted before the condition is available, starting speculation.
\end{itemize}

The semantic state $\langle \sigma, \x{spc}, E, B \rangle$ is composed of the architectural state $\sigma$, speculative program counter $\x{spc}$ that points to the next instruction to issue, reorder buffer $E$ and branch predictor state $B$.
We model the reorder buffer $E$ as a sequence of triples of the form $\langle \x{pc}, v_\bot, b \rangle$, where $\x{pc}$ is the address of the instruction in the reorder buffer, $v_\bot$ is the result value of the instruction if it was evaluated out-of-order, and $b \in \Bools$ is the branch prediction for branch instructions, ignored for non-branches.

The operational semantics for speculative execution is split into three groups of rules. The \emph{issue} rules (Figure~\ref{fig:specissue}), which for instructions that do not speculate, insert an entry into the reorder buffer for that instruction. For the $\IBranch$ instruction, the issue rule ignores the condition and instead uses the branch predictor to determine the new program counter, recording the branch prediction in the reorder buffer entry (rule S-Branch-Issue). The \emph{execute} rules (Figure~\ref{fig:specexec}) evaluate instructions with available dependencies. We only have execute rules for instructions that produce values, the available values in registers and memory are computed by auxiliary reorder buffer lookup predicates in Figure~\ref{fig:specaux}. The \emph{commit} rules (Figure~\ref{fig:speccommit}) update the architectural state with the result of the oldest instruction in the reorder buffer. Since our architectural semantics is determistic, we just lift the architectural semantics rules to the microarchitectural rules for all the instructions that cannot mis-speculate (S-NS-Commit). When committing the $\IBranch$ instruction, we flush the reorder buffer if the speculated condition turned out to be wrong (rule S-Predict-Fail).

\begin{figure}[th]
  \begin{align*}
    \srule{\mathrm{nospec}(\rho[\x{spc}])}
          {\sj{\rho}{\langle \sigma, \x{spc}, E, B \rangle}
                    {\langle \sigma, \x{spc} + 1, E \concat [\langle \x{spc}, \bot, \bot \rangle],
                     B \rangle}}
          {S-Seq-Issue}
    \\[1em]
    \srule{\begin{array}{r c l}
            \rho[\x{spc}] & = & \IBranch \hsp r_c \hsp d \\
             b & = & \predict(B, \x{spc}) \\
             \x{spc}' & = &\left\{
             \begin{array}{l l l}
               \x{spc} + d & \textrm{if} \quad  b = \True \\
               \x{spc} + 1 & \textrm{otherwise}
             \end{array}
           \right.
           \end{array}
          }
          {\sj{\rho}{\langle \sigma, \x{spc}, E, B \rangle}
                 {\langle \sigma, \x{spc}', E \concat [\langle \x{spc}, \bot, b \rangle],
                 B' \rangle}}
          {S-Branch-Issue}
  \end{align*}
  \noindent where non-speculating instruction predicate is defined by
  \begin{align*}
    \mathrm{nospec}(i) = &\left\{
             \begin{array}{l l l}
               \True & \textrm{if} &
                  i = \IConst \hsp r_d \hsp k\\
               & \textrm{or} & i = \ILoad  \hsp r_d \hsp [r_a + \x{k}]\\
               & \textrm{or} & i = \IStore \hsp [r_a + \x{k}] \hsp r_v\\
               & \textrm{or} & i = \IBinop \hsp \x{op} \hsp r_d \hsp r_a \hsp r_b \\
               \False & \multicolumn{2}{l}{\textrm{otherwise}} \\
             \end{array}
             \right.
  \end{align*}

  \caption{Speculative semantics - issue instructions.}
  \label{fig:specissue}
\end{figure}

\begin{figure}[ht]
  \begin{align*}
    \srule{\begin{array}{c}
             \rho[\x{spc}'] = \ILoad  \hsp r_d \hsp [r_a + \x{k}] \\
             \lookupR(\rho, R, E, r_a) = v_a \qquad
             \lookupM(\rho, R, M, E, v_a + k) = v
           \end{array}}
          {\sj{\rho}{\langle \sigma, \x{spc}, E \concat [\langle \x{spc}', \bot, \bot \rangle]
                 \concat E', B \rangle}
                 {\langle \sigma, \x{spc}, E \concat [\langle \x{spc}', v, \bot \rangle]
                 \concat E', B \rangle}}
          {S-Load-Execute}
    \\[1em]
    \srule{\begin{array}{c}
             \rho[\x{spc}'] = \IBinop \hsp \x{op} \hsp r_d \hsp r_a \hsp r_b \\
             \lookupR(\rho, R, E, r_a) = v_a \qquad
             \lookupR(\rho, R, E, r_b) = v_b \\
             v = \x{op}(v_a, v_b)
           \end{array}}
          {\sj{\rho}{\langle \sigma, \x{spc}, E \concat [\langle \x{spc}', \bot, \bot \rangle]
                 \concat E', B \rangle}
                 {\langle \sigma, \x{spc}, E \concat [\langle \x{spc}', v, \bot \rangle]
                 \concat E', B \rangle}}
          {S-Op-Execute}
  \end{align*}
  \caption{Speculative semantics - execute instructions.}
  \label{fig:specexec}
\end{figure}

\begin{figure}[ht]
  \begin{align*}
    \srule{\nospec(\rho[\x{pc}]) \qquad \nj{\rho}{\sigma}{\sigma'}}
          {\sj{\rho}{\langle \sigma, \x{spc},
                       \langle \x{pc}, v_\bot, b \rangle \concat E,
                       B \rangle}
                 {\langle \sigma', \x{spc}, E, B \rangle}}
          {S-NS-Commit}
    \\[1em]
    \srule{\begin{array}{r c l}
              \rho[\x{pc}] & = & \IBranch \hsp r_c \hsp d \\
              \langle b, \x{pc}' \rangle & = & \left\{
              \begin{array}{l l l}
                \langle \True, \x{pc} + d \rangle & \mathrm{if} & R[r_c] \neq 0 \\
                \langle \False, \x{pc} + 1 \rangle & \mathrm{if} & R[r_c] = 0
              \end{array}
              \right. \\
             B' & = & \x{BP_{\Max{B}}}(B, b)
            \end{array}
          }
          {\sj{\rho}{\langle \langle \x{pc}, R, M \rangle, \x{spc}, [\langle \x{pc}, \bot, b \rangle]
                 \concat E, B \rangle}
                 {\langle \langle \x{pc}', R, M \rangle, \x{spc}, E, B' \rangle}}
          {S-Predict-Success}
    \\[1em]
    \srule{\begin{array}{r c l}
              \rho[\x{pc}] & = & \IBranch \hsp r_c \hsp d \\
              \langle b, \x{pc}' \rangle & = & \left\{
              \begin{array}{l l l}
                \langle \False, \x{pc} + d \rangle & \mathrm{if} & R[r_c] \neq 0 \\
                \langle \True, \x{pc} + 1 \rangle & \mathrm{if} & R[r_c] = 0
              \end{array}
              \right. \\
             B' & = & \x{BP_{\Max{B}}}(B, \invert(b))
            \end{array}
          }
          {\sj{\rho}{\langle \langle \x{pc}, R, M \rangle, \x{spc}, [\langle \x{pc}, \bot, b\rangle]
                 \concat E, B \rangle}
                 {\langle \langle \x{pc}', R, M \rangle, \x{spc}, [], B' \rangle}}
          {S-Predict-Fail}
  \end{align*}
  \caption{Speculative semantics - commit instructions.}
  \label{fig:speccommit}
\end{figure}

\begin{figure}[ht]
  \begin{align*}
    \srule{\rho[\x{spc}] = \ILoad \hsp r_d \hsp [r_a + \x{k}]}
          {\lookupR(\rho, R, E \concat [\x{spc}, v^\bot, \bot], r_d) = v^\bot}
          {S-LookupR-Load}
    \\[1em]
    \srule{\rho[\x{spc}] = \IBinop \hsp \x{op} \hsp r_d \hsp r_a \hsp r_b}
          {\lookupR(\rho, R, E \concat [\x{spc}, v^\bot, \bot], r_d) = v^\bot}
          {S-LookupR-Op}
    \\[1em]
    % This is sloppy quantification, perhaps fix or use disjunction here?
    \srule{\begin{array}{c}
             \rho[\x{spc}] \neq \ILoad \hsp \_ \hsp \_ \quad
             \rho[\x{spc}] \neq  \IBinop \hsp \_ \hsp \_ \hsp \_ \hsp \_ \\
             \lookupR(\rho, R, E, r_v) = v^\bot
           \end{array}}
          {\lookupR(\rho, R, E \concat [\x{spc}, v_*^\bot, \bot], r_v) = v^\bot}
          {S-LookupR-Other}
    \\[1em]
    \srule{ }
          {\lookupR(\rho, R, [], r) = R[r]}
          {S-LookupR-Base}
    \\[1em]
    \srule{\begin{array}{r c l}
             \rho[\x{spc}] & = & \IStore \hsp [r_a + \x{k}] \hsp r_v \\
             \lookupR(\rho, R, E, r_a) & = & v_a \\
             \lookupR(\rho, R, M, E, r_v) & = & v
           \end{array}}
          {\lookupM(\rho, R, M, E \concat [\x{spc}, \bot, \bot], v_a + k) = v}
          {S-LookupM-Store-Value}
    \\[1em]
    \srule{\begin{array}{c}
             \rho[\x{spc}] = \IStore \hsp [r_a + \x{k}] \hsp r_v \\
             \lookupR(\rho, R, E, r_a) = \bot
           \end{array}}
          {\lookupM(\rho, R, M, E \concat [\x{spc}, \bot, \bot], v_a) = \bot}
          {S-LookupM-Store-Unknown}
    \\[1em]
    \srule{\begin{array}{r c l}
      \rho[\x{spc}] & = & \IStore \hsp [r_a + \x{k}] \hsp r_v \\
             \lookupR(\rho, R, E, r_a) + k  & \neq & v_a \\
             \lookupM(\rho, R, M, E, v_a) & = & v^\bot
           \end{array}}
          {\lookupM(\rho, R, M, E \concat [\x{spc}, \bot, \bot], v_a) = v^\bot}
          {S-LookupM-Store-NoAlias}
    \\[2em]
    \srule{\begin{array}{c}
             \rho[\x{spc}] \neq \IStore \hsp \_ \hsp  \_ \\
             \lookupM(\rho, R, M, E, v) = v^\bot
           \end{array}}
          {\lookupM(\rho, R, M, E \concat [\x{spc}, \bot, \bot], v_a) = v^\bot}
          {S-LookupM-Other}
    \\[1em]
    \srule{ }
          {\lookupM(\rho, M, [], v) = M[v]}
          {S-LookupM-Base}
  \end{align*}
  \caption{Auxiliary predicates}
  \label{fig:specaux}
\end{figure}

It is useful to observe that the branch speculative semantics coincides with the architectural semantics until the first misspeculation in the reorder buffer. To make the notions of state agreement and misspeculation precise, we will say that program $\rho$'s architectural state $\sigma = \langle \x{pc}, R, M \rangle$ agrees with its microarchitectural state $\langle \langle \x{pc}', R', M' \rangle, \x{spc}, \{e_i\}^n_{i=0}, B \rangle$ at depth $m \leq n$, if
\begin{itemize}
  \item for all registers $r$, either $\lookupR(\rho, R', \{e_i\}^m_{i=0}, r)$ is bottom or it equals to $R[r]$,
  \item for all memory locations $l$, value $\lookupM(\rho, R', M', \{e_i\}^m_{i=0}, r)$ equals to either bottom or $M[l]$,
  \item The program counter $\x{pc}$ is equal to the program counter of $e_{m+1}$ if $m < n$ or $\x{pc} = \x{spc}$.
\end{itemize}

For program $\rho$, we define the state after applying reorder buffer entries $\{e_i\}^m_{i=0}$, written $\mathrm{apply}(\rho, \sigma, \{e_i\}^m_{i=0})$ to be the unique state that is reached after $m$ transitions, i.e., $\rho, \sigma \longrightarrow_m \mathrm{apply}(\rho, \sigma, \{e_i\}^m_{i=0})$. Note that we do not have to decode the buffer because our architectural semantics is deterministic. For a non-determinstic architecture, we would have to replay the instructions contained in the reorder buffer.

We will say that program $\rho$'s microarchitectural state \emph{mis-speculates at reorder buffer depth $m$} if $m$-th entry in the reorder buffer is a branch with prediction $b$ and evaluating the condition by applying the first $m$ instructions from the reorder buffer on the architectural state disagrees with $b$. More precisely, we say $\langle \langle \x{pc}', R', M' \rangle, \x{spc}, \{\langle \x{pc}_i, v_i, b_i \rangle \}^n_{i=0}, B \rangle$ mis-speculates in reorder buffer depth $m \leq n$ if $\rho[\x{pc}_m] = \IBranch \hsp r_c \hsp \x{d}$ and for $\mathrm{apply}(\rho, \langle \x{pc}', R', M' \rangle, \{\langle \x{pc}_i, v_i, b_i \rangle \}^n_{i=0}) = \langle \x{pc}'_m, R'_m, M'_m \rangle$, we have either $R'_m[r_c] = 0$ and $b_m = \True$ or $R'_m[r_c] \neq 0$ and $b_m = \False$.

\begin{theorem}\label{thm:spec-arch-agree}
Given a program $\rho$'s microarchitectural state $\mu = \langle \sigma, \x{spc}, \{e_i\}^n_{i=0}, B \rangle$ reachable from a state with empty reorder buffer, either the state $\mathrm{apply}(\rho, \sigma, \{e_i\}^m_{i=0})$ agrees with $\mu$ at all depths $m \leq n$ or $\mu$ mis-speculates at depth $k$ and $\mathrm{apply}(\rho, \sigma, \{e_i\}^m_{i=0})$ agrees with $\mu$ at depths $m \leq k$.
\end{theorem}

% TODO(titzer/jarin) - this seemed not to be used or fleshed out, so dropped it for space.
% \subsubsection{Memory Disambiguator State}
%
% Memory disambiguator speculatively bypasses stores in reorder buffer when executing loads. We model this by allowing reorder buffer memory lookup rule (cf. Figure~\ref{fig:specaux}) ignore stores.
% \begin{align*}
%     \srule{\begin{array}{c}
%              \lookupM(\rho, R, M, E, v) = v^\bot
%            \end{array}}
%           {\lookupM(\rho, R, M, E \concat [\x{spc}, \bot, \bot], v_a) = v^\bot}
%           {S-LookupM-Other}
%   \end{align*}

\subsubsection{Timer}

To implement a counter that counts \MuArchal steps, we add \MuState $\x{T} \in \Nats$ and extend the semantics with a meta-rule that increments the count for every step of evaluation. Reading the timer is then accomplished with the straightforward rule.

% TODO(titzer): do we want to start adding alpha and mu over the semantics arrows?
{\small
  \begin{align*}
    \srule{\rho, \langle pc, R, M \rangle \xlongrightarrow{\alpha} \langle pc', R', M' \rangle \hspw T' = T+1}
          {\nj{\rho}{\langle \x{pc}, R, M, T \rangle}
                 {\langle \x{pc'}, R', M', T' \rangle}}
          {Tick-Rule}
    \\[1em]
    \srule{\rho[\x{pc}] = \ITimer \hsp r_d \hspw T' = T + 1}
          {\nj{\rho}{\langle \x{pc}, R, M, T \rangle}
                 {\langle pc + 1, \supdate{R}{r_d}{T}, M, T' \rangle}}
          {Timer-Read}
    \end{align*}
}

\subsection{Composing \MuArchal extensions}\label{sec:composition}

As we've seen in the previous sections, a number of extensions are possible to model \MuArchal mechanisms by adding state and additional evaluation rules.
To create a complete model of a \MuArch, we can \emph{compose} these extensions into a larger model that contains state for caches, branch prediction, indirect branch prediction, out-of-order execution, and a timer.
As most of these extensions have orthogonal state, this is generally straightforward (though perhaps tedious).

\subsection{Cache-based side-channels}  \label{sec:side-channels}

Real CPU caches copy data in physical memory, include virtual or physical tags, and include \MuState for tracking dirtiness and implementing coherency protocols across multiple cache levels spread over many cores and processors. Our model is thus very simplified, since these fine details are not needed for our purposes.

\subsubsection{Encoding information in cachedness}

Caches are typically measured by their capacity to store program data, yet we recognize they have a \emph{meta-information capacity}, $C_\x{info}$ that represents the information they store about program history. This capacity depends on the exact details of the \MuState, such as the number of levels, replacement policy, inclusiveness, etc. In our model, $C_\x{info} = \Max{C} log\hspace{2pt} |M|$\footnote{Note that despite first intuition, we cannot actually store a bit per address, since only \Max{C} bits can simultaneously be \One.} where $\Max{C}$ and $|M|$ are respectively the cache capacity and the maximum memory address.

We need an encoding scheme to store information in the cachedness of memory. The fact that all memory accesses may alter the state of the cache means it must be robust against:
\begin{itemize}
  \item \emph{incidental} accesses that occur between storing information in the cache and its retrieval
  \item accesses by the implementation of the encoding scheme itself, particularly retrieval
  \item \emph{noise} caused by interrupt handlers and concurrent processes sharing the same cache
  \end{itemize}

The decoding problem restricts the amount of meta-information we can store, so in practice $C_\x{info}$ is not achievable. In practice, an attack may only require leaking a single bit or byte at a time. Two schemes are:

\begin{itemize}
\item \textbf{Direct-mapped bits}. For a small number $b < \Max{C}$ of bits, we choose addresses $a_0 \ldots a_{b-1}$. We store a \One for bit $i$ by accessing $a_i$, bringing it into the cache. We store a \Zero for bit $i$ by \emph{evicting} $a_i$ from the cache\footnote{We can evict a line by filling the cache with \emph{other} addresses that alias it, called an \emph{eviction set}.}. A program can read bit $i$ by measuring the access time to $a_i$ using the \ITimer instruction, determining \One if fast, \Zero if slow.
\item \textbf{Indexed values}. For a small number $b < log \Max{C}$ of bits, we use a larger number of addresses $a_0 \ldots a_{2^b-1}$. To store the value $v = B_{b-1} \ldots B_1 B_0$, we access address $a_v$. To read a value from the cache, we probe lines $a_0 \ldots a_{2^b-1}$. The fast address $a_v$ indicates the value $v$.
\end{itemize}

Both schemes have advantages and disadvantages. Direct-mapped is space-efficient but requires $b$ accesses to write $b$ bits. It is less robust to information loss, since each stored \One bit is at risk of being flipped to \Zero by eviction through incidental accesses. Indexed-value has the advantage that one access can write $b$ bits, but it requires $2^b$ accesses to read $b$ bits. It is also more robust to errors, since the one cached memory address which carries information is less likely to be evicted by accident. In our work on real hardware, we typically used the indexed-value scheme with $4 \leq b \leq 8$, both because it was easier to leak information from speculation this way, and it was more robust to errors and noise.

\subsection{Vulnerabilities by the Numbers}\label{sec:variants}

\subsubsection{Variant 1: Speculative Safety Check Bypass}

\begin{figure}
{\tiny
\centering
\begin{subfigure}{0.3\textwidth} 
  \centering
  \vspace{4ex}
  \begin{lstlisting}
function vulnerable(index) {
  if (index >= array.length)
    return 0;
  return timing[array[index]];
}

vulnerable(1);  // Train
vulnerable(3);  // Attack
  \end{lstlisting}
  \caption{Vulnerable code}
  \label{subfig:var1code}
\end{subfigure}
~
\begin{subfigure}{0.5\textwidth}  
  \centering
  \begin{lstlisting}
 0: load r1 [#array - 1]   ; load array length
 1: binop[ge] r2 r0 r1     ; index >= array length
 2: branch r2 @oob         ; (vulnerable) bounds check
 3: load r0 [r0 + #array]  ; load [array + index]
 4: load r0 [r0 + #timing] ; load [timing + val]
 5: jump rip               ; return
@oob:
 6: const r0 #0            ; out of bounds => 0
 7: jump rip               ; return
  \end{lstlisting}
  \caption{Vulnerable machine code}
  \label{subfig:var1machinecode}
\end{subfigure}

\hspace{-5ex}
\begin{subfigure}{0.3\textwidth}
  \centering
  \vspace{-20ex}
  \begin{tikzpicture}[]
  \matrix(memory)[column sep=0mm, ampersand replacement=\&] {
    \node(){0}; \& \node(memTimingStart)[draw]{0}; \\ 
    \node(){1}; \& \node()[draw]{1}; \\ 
    \node(){2}; \& \node()[draw]{2}; \\
    \node(){3}; \& \node(memTimingEnd)[draw]{3}; \\ 
    \node(){4}; \& \node(memLen)[draw]{2}; \\
    \node(){5}; \& \node(memArrayStart)[draw]{3}; \\
    \node(){6}; \& \node(memArrayEnd)[draw]{1}; \\
    \node(){7}; \& \node(memSecret)[draw]{2}; \\
  };
  \memLabel{timing}{memTimingStart}{memTimingEnd}
  \memLabel{len}{memLen}{memLen}
  \memLabel{array}{memArrayStart}{memArrayEnd}
  \memLabel{secret}{memSecret}{memSecret}
  \end{tikzpicture}
\caption{Memory layout}
\end{subfigure}
~
\begin{subfigure}[t]{0.6\textwidth}
\mbox{
  \realArchState{1}{0}{0}{0}
  \uArchState{1}{0}{0}{0}
  \cacheState{}{}
  \drawCpu
~
  \realArchState{1}{2}{0}{3}
  \uArchState{1}{2}{0}{3}
  \cacheState{\emph{not-taken}}{}
  \drawCpu
~
  \realArchState{3}{2}{0}{5}
  \uArchState{3}{2}{0}{5}
  \cacheState{\emph{not-taken}}{\emph{timing} $+$ 3}
  \drawCpu
}
\caption{Training the predictor}
\label{subfig:var1train}
\end{subfigure}

\hspace{-15ex}
\begin{subfigure}[t]{1\textwidth}
\mbox{
  \realArchState{3}{0}{0}{0}
  \uArchState{3}{0}{0}{0}
  \cacheState{\emph{not-taken}}{}
  \drawCpu
~
  \realArchState{3}{0}{0}{0}
  \uArchState{3}{?}{?}{3}
  \cacheState{\emph{not-taken}}{}
  \drawCpu
~
  \realArchState{3}{0}{0}{0}
  \uArchState{2}{?}{?}{5}
  \cacheState{\emph{not-taken}}{\emph{timing} $+$ 2}
  \drawCpu
~
  \realArchState{3}{2}{1}{2}
  \uArchState{3}{2}{1}{2}
  \cacheState{\emph{?}}{\emph{timing} $+$ 2}
  \drawCpu
~
  \realArchState{0}{2}{1}{7}
  \uArchState{0}{2}{1}{2}
  \cacheState{\emph{taken}}{\emph{timing} $+$ 2}
  \drawCpu
}
\caption{Performing the attack}
\label{subfig:var1attack}
\end{subfigure}
}  % /tiny
\caption{Illustration of a variant 1 attack}
\label{fig:var1}
\end{figure}
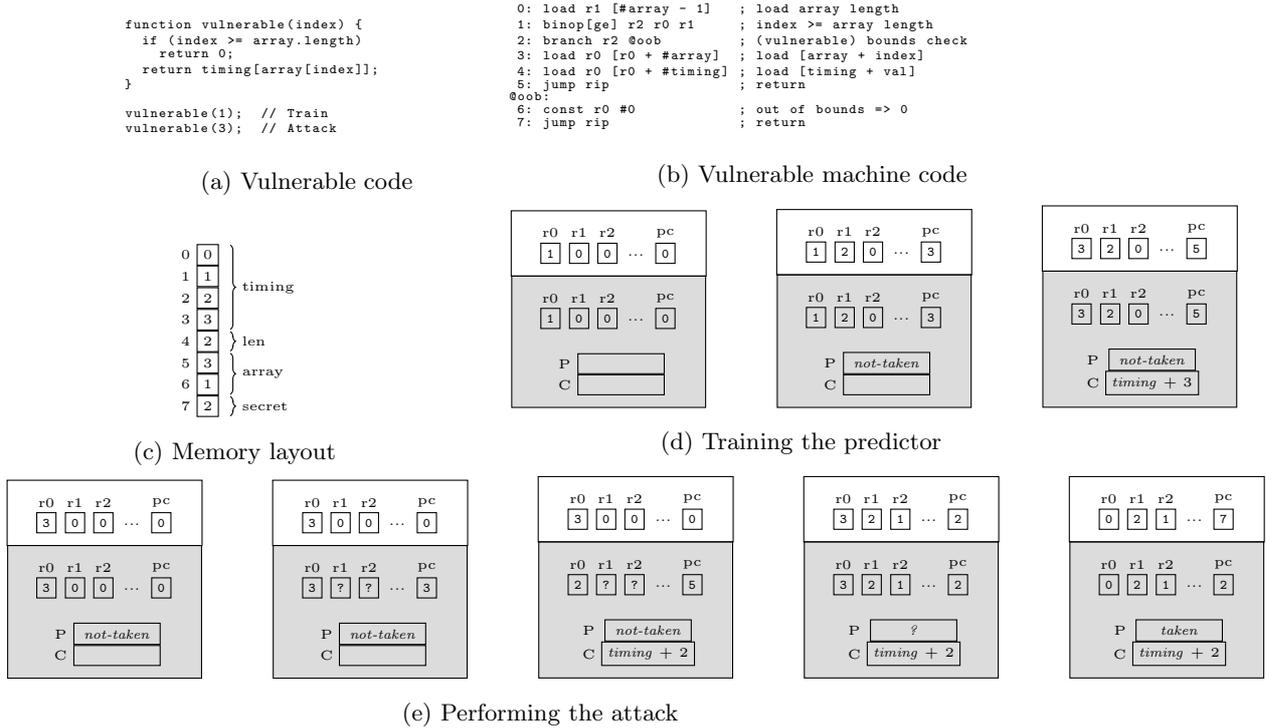

Programs often contain branches that implement safety checks to prevent unsafe runtime behavior such as accessing outside the bounds of allocated memory or accessing an object of an incorrect type. Depending on the programming language semantics, an out-of-bounds access might result in a language-level exception being thrown or a sentinel value like \texttt{undefined} or \Zero being returned. Most programs are well-behaved, so safety checks normally pass. When executing such programs, CPU branch predictors quickly learn to predict these branches. The uncommon failed safety check will result in the processor's normal recovery mechanism for incorrectly predicted branches rolling back architectural state to before the mispredicted branch and instead executing the proper architectural path. However, as we have seen in our model, CPUs do not generally rollback the \MuState changes, so changes to caches are not undone. Therein lies our first vulnerability.

An attacker can construct a program that trains the CPU's branch predictor to assume that safety checks normally pass, and then intentionally triggers a misprediction that results in speculatively executing code where \emph{safety conditions normally established by branches do not hold}. To date, we've all assumed that this was innocent because architectural state rollback would not allow an attacker to make use of any information accessed in misspeculated executions. However, a careful attacker can exfiltrate information from speculative execution through \MuState.

The example in Figure~\ref{fig:var1} shows a program that attacks bounds checks.
The routine \texttt{vulnerable} accepts an integer index argument.
Compiled code loads the array length and then checks the index against this length.
If the index is in-bounds, it loads the array element and then uses that value as an index into a second array, \texttt{timing}.
If the index is out-of-bounds, the code simply returns \Zero.
Clearly, no execution of this program should access outside the bounds of the array, as the load is guarded by a bounds check.
However, the attacker trains the branch predictor to assume \emph{not-taken} (Figure~\ref{subfig:var1train}, with \MuState shown in light gray). After training, the attacker crafts a special out-of-bounds access to cause the CPU to mis-speculate and load the out-of-bounds memory location (Figure~\ref{subfig:var1attack}). Even though the CPU rolls-back the speculative architectural state and executes \texttt{@oob}, the value at \texttt{secret} has already been encoded into \MuState as the cachedness of the address \texttt{timing}$+\x{secret}$. The attacker decodes the secret data using the techniques in Section \ref{sec:side-channels}.

\begin{figure}[t]
{\tiny
\centering
\begin{subfigure}{0.4\textwidth} 
  \centering
  \begin{lstlisting}
class A { int* a; };
function A.virtualFunc() { 
  return timing[*(this.a)]
}
class B { int b; };
function B.virtualFunc() { 
  return this.b
}
function vulnerable(obj) {
  return obj.virtualFunc();
}

vulnerable(new A(&val)); // Train
int addr = <craft secret address>;
vulnerable(new B(addr);  // Attack
  \end{lstlisting}
  \caption{Vulnerable code}
  \label{subfig:var12:code}
\end{subfigure}
~
\begin{subfigure}{0.5\textwidth}  
  \centering
  \vspace{8ex}
  \begin{lstlisting}
 0: load r1 [r0 + #0]       ; load obj.virtualFunc
 1: jump r1                 ; (vulnerable) indirect jump

@A.virtualFunc:
 2: load r2 [r0 + #1]       ; load this.a
 3: load r1 [r2]            ; load *(this.a)
 4: load r0 [r1 + #timing]  ; load [timing + val]
 5: jump r15                ; return

@B.virtualFunc:
 6: load r0 [r0 + #1]       ; load this.b
 7: jump r15                ; return
  \end{lstlisting}
  \caption{Vulnerable machine code}
\end{subfigure}

\hspace{-5ex}
\begin{subfigure}{0.3\textwidth}
  \centering
  \vspace{-22ex}
  \begin{tikzpicture}[]
  \matrix(memory)[column sep=0mm, ampersand replacement=\&] {
    \node(){0}; \& \node(memTimingStart)[draw]{0}; \\ 
    \node(){1}; \& \node()[draw]{1}; \\ 
    \node(){2}; \& \node()[draw]{2}; \\
    \node(){3}; \& \node(memTimingEnd)[draw]{3}; \\ 
    \node(){4}; \& \node(AStart)[draw]{2}; \\
    \node(){5}; \& \node(AEnd)[draw]{6}; \\
    \node(){6}; \& \node(AInt)[draw]{3}; \\
    \node(){7}; \& \node(BStart)[draw]{6}; \\
    \node(){8}; \& \node(BEnd)[draw]{9}; \\
    \node(){9}; \& \node(secret)[draw]{2}; \\
  };
  \memLabel{timing}{memTimingStart}{memTimingEnd}
  \memLabel{A.virtualFunc}{AStart}{AStart}
  \memLabel{A.a}{AEnd}{AEnd}
  \memLabel{integer pointed to by A.a}{AInt}{AInt}
  \memLabel{B.virtualFunc}{BStart}{BStart}
  \memLabel{B.b (address of secret)}{BEnd}{BEnd}
  \memLabel{secret}{secret}{secret}
  \end{tikzpicture}
\caption{Memory layout}
\end{subfigure}
~
\begin{subfigure}[t]{0.6\textwidth}
\mbox{
  \realArchState{4}{0}{0}{0}
  \uArchState{4}{0}{0}{0}
  \cacheState{}{}
  \drawCpu
~
  \realArchState{4}{2}{0}{2}
  \uArchState{4}{2}{0}{2}
  \cacheState{\texttt{@1 -> @2}}{}
  \drawCpu
~
  \realArchState{3}{3}{6}{5}
  \uArchState{3}{3}{6}{5}
  \cacheState{\texttt{@1 -> @2}}{\emph{timing} $+$ 3}
  \drawCpu
}
\vspace{-0.75ex}
\caption{Training the predictor}
\label{subfig:var12train}
\end{subfigure}

\hspace{-15ex}
\begin{subfigure}[t]{1\textwidth}
\mbox{
  \realArchState{7}{0}{0}{0}
  \uArchState{7}{0}{0}{0}
  \cacheState{\texttt{@1 -> @2}}{}
  \drawCpu
~
  \realArchState{7}{0}{0}{0}
  \uArchState{7}{?}{0}{1}
  \cacheState{\texttt{@1 -> @2}}{}
  \drawCpu
~
  \realArchState{7}{0}{0}{0}
  \uArchState{2}{2}{9}{5}
  \cacheState{\texttt{@1 -> @2}}{\emph{timing} $+$ 2}
  \drawCpu
~
  \realArchState{7}{6}{0}{1}
  \uArchState{7}{6}{0}{1}
  \cacheState{\texttt{@1 -> ?}}{\emph{timing} $+$ 2}
  \drawCpu
~
  \realArchState{1}{6}{0}{7}
  \uArchState{1}{6}{0}{7}
  \cacheState{\texttt{@1 -> @6}}{\emph{timing} $+$ 2}
  \drawCpu
}
\caption{Performing the attack}
\label{subfig:var12attack}
\end{subfigure}
}  % /tiny
\caption{Illustration of a indirect branch variant 1 attack}
\label{fig:var12}
\end{figure}

Another vulnerability is exposed by indirect jumps, which are typically used by programs to implement type-dispatched behavior, numerically-indexed \texttt{switch} statements and threaded bytecode interpreters.
Often the construction of objects in memory enforces implicit safety properties.
For example, a typical implementation of classes stores a header word before each object in memory which points to metadata for type A, such as a virtual dispatch table or \emph{vtable}, if and only if the object is of type A in the source program.
An indirect jump through the virtual dispatch table allows the target of the indirect jump to assume, without checking, that receiver objects are of the proper type.

The example shown in Figure~\ref{fig:var12} shows that indirect branch speculation violates this assumption. The routine \texttt{vulnerable} accepts a single argument which is the address of an object. It assumes that the first memory cell of the object contains a pointer to code that implements a virtual function \texttt{virtualFunc}. If the routine is repeatedly called with objects of type \texttt{A}, the indirect branch predictor will predict that the indirect call always jumps to \texttt{@A.virtualFunc} (Figure~\ref{subfig:var12train}). When the routine is then called with an object of type \texttt{B}, the CPU mispredicts and speculatively jumps to \texttt{@A.virtualFunc}, but with an argument of type \texttt{B} (Figure~\ref{subfig:var12attack}). Objects of type \texttt{A} contain a pointer in the second cell, while objects of type \texttt{B} contain an integer in the second cell. Thus when \texttt{@A.virtualFunc} is executed speculatively on an object of type \texttt{B}, the CPU will confuse an integer field  as a pointer, speculatively loading from this \emph{attacker-crafted pointer}. The attacker ex-filtrates the value again through \MuState of the cache.

\subsubsection{Variant 2: Speculative Target Misreconstruction}

In variant 1 we've shown that indirect jump prediction can be exploited to bypass the implicit type checks that are part of a typical language's virtual dispatch mechanism. As it turns out, the branch target buffer on most CPUs are \emph{approximate} in order to save space. For example, Intel 64-bit CPUs only store the low-order 32 bits of the \emph{from} address (the address of the indirect jump) and the low-order 32 bits of the \emph{relative target} address (the predicted address). Upon lookup, the predictor ignores the upper 32 bits of the \emph{from} address and reuses a prediction \emph{for an aliased from address}. This allows an attacker to train a target indirect branch to speculatively jump to any address within a 4GiB range \emph{without ever executing the victim branch}. 

This is particularly bad, because the attacker can create \emph{speculative indirect jumps to anywhere}, i.e., control flow that cannot possibly exist in the original code, such as jumping into the middle of arbitrary machine code that simply happens to be a leaking gadget. That means an attacker may not even need to craft an instruction sequence, but find an extant instruction sequence in the victim's code, similar to return-oriented programming. This can even work across processes. \cite{Kocher2018spectre} found that the branch target buffer on Intel chips is shared across hyperthreads on the same core, allowing one process to inject predictions into another. As a mitigation for this attack, a subsequent microcode update from Intel disabled this sharing~\cite{IntelMicrocodeV2}.

Appendix~\ref{app:var2} provides an illustrative example of a variant 2 attack.

\subsubsection{Variant 3: Speculative Hardware Permission Check Bypass}

In addition to programmatic safety checks to prevent unsafe runtime behavior, the hardware itself provides certain guarantees via implicit permission checks. User programs should not be able to access unmapped virtual memory addresses, write to read-only memory~\cite{Variant1.1}, or read from kernel memory. Such attempts should result in a \emph{faults}.
Some CPUs seem to check for a fault too late, effectively speculating through a hardware permission check.
This depends on the specific details of a CPU's trap mechanism of course; e.g. faulting at retirement is too late if the processor has already accessed the memory and supplied its value to dependent instructions, which leaked the value into \MuState. Lipp et al.~\cite{Lipp2018meltdown}, describe a variant 3 attack that enables leakage of data in kernel memory to a userspace process.

These hardware checks are sometimes used by the CPU as a general mechanism to interrupt the program and transfer control to the kernel. It has been discovered that Intel CPUs use it to implement an optimization called Lazy FPU state restore~\cite{LazyFPU}. Upon context switch between two applications, microprocessors can choose to lazily restore floating-pointer registers of the arriving process. If the program reads from an unrestored register, it receives a stale value (from a previous process) and continues executing speculatively, faulting at retirement. Again, this is too late, since the process might leak the stale value in \MuState.

\subsubsection{Variant 4: Speculative Aliasing Confusion} \label{sec:variant4}

Since memory is often the bottleneck in many programs,  modern CPUs utilize not only caching but dynamic alias analysis known as \emph{memory disambiguation}. When executing a store, the CPU uses a predictor to determine which, if any, subsequent loads will depend on the store. If the prediction is \emph{no-alias}, the CPU may speculatively execute a later load before the store. If the prediction turns out to be incorrect and the store address and load address are in fact aliases, this will be detected when instructions are being retired in program order, and the load will be aborted and re-executed. This, too, represents a vulnerability, since loads that are speculatively executed out of order observe stale values from memory. 

Bypassing stores is only one way a memory disambiguator can speculatively accelerate loads~\cite{SpeculativeForwarding}. As long as violations are detected and repaired before retirement, other aggressive forwarding strategies could be implemented. If the memory disambiguator learns that a load typically aliases a store, it could speculatively forward the value even if the source address for the load is not yet known. Similarly the disambiguator could learn that two consecutive loads typically load from the same address, and inject the result from the first load into the second. Like in variant 2, predictions made by the memory disambiguator are subject to implementation details.
%REDACTED(PSF disclosure) Even if addresses of loads and stores are already fully computed, the disambiguator could forward based only on partial matches of the addresses.

Appendix~\ref{app:var4} provides an illustrative example of a variant 4 attack.

\subsection{The Universal Read Gadget}

By design, misspeculated execution has no architectural side-effects.
That means that no speculative execution vulnerability can alter the architectural state; they are limited to read-only access. However, as we have seen, the four variants we have outlined bypass normal software safety checks and the assumption of language type safety, allowing even a well-typed program to read inaccessible memory.
The most general form of out-of-bounds read, a routine that can read all of addressable memory, we term the \emph{universal read gadget}.

Our work on Spectre has led us to the disheartening conclusion:

\begin{itemize}
\item \textbf{Pervasive availability of the Universal read gadget.} For most programming languages L with a timer, speculative vulnerabilities on most of today's CPUs allow the construction of a well-typed procedure\newline
\newline 
 \texttt{read(address: int, bit: int) $\rightarrow$ bit}\newline
\newline
  that uses a \MuArchal side-channel to read the \texttt{bit} of the contents of the process memory at the given \texttt{address}.
\end{itemize}

The universal read gadget is not necessarily a straightforward construction. It requires detailed knowledge of the \MuArchal characteristics of the CPU and knowledge of the language implementation, whether that be a static compiler or a virtual machine. Additionally, the gadget might have particularly unusual performance and concurrency characteristics:
\begin{itemize}
\item if the timer is low resolution, the gadget requires amplification
\item the gadget may require training \MuArchal predictors in a complex warmup phase
\item the gadget may fail probabilistically due to noise from interrupts, frequency scaling, or predictor algorithms with hidden state, and thus requires repeated attempts
\end{itemize}

What characteristics of a programming language make it exploitable on today's modern hardware?
As we have seen, access to a timer, no matter the resolution, leaks \MuArchal information.
We point out several language features whose typical implementations may be vulnerable to Spectre.
In these we found that a key to constructing the universal read gadget was speculative \emph{pointer crafting}, whereby an attacker exploits speculation to trick the implementation into interpreting attacker-controlled input as a machine-level pointer, feeding this pointer into a (normally innocent, but speculatively dangerous) load to achieve the universal read gadget.

Source-language features we have determined to be vulnerable include:

\begin{enumerate}
\item \textbf{Indexed data structures with dynamic bounds checks.} As seen in our example variant 1 vulnerability, bounds checks inserted into code by either static or dynamic compiler can be trained to be mispredicted, causing speculative out-of-bounds accesses.
\item \textbf{Differential data structure shapes.} Data structures of different types have different layouts in memory. Code manipulating objects of different shapes, e.g. objects of two different classes, can be trained to execute with improper object layouts, causing type confusion leading to pointer crafting~\cite{TypeConfusion}.
\item \textbf{Dynamically shaped structures.} In languages that have dynamically typed objects where the underlying storage may change shape, mispredicted shape checks may allow an attacker to read out of the bounds of an object's shape into nearby objects, which can contain attacker-controlled values.
\item \textbf{Variadic arguments to functions.} Like other dynamic data structures, both variant 1 and variant 4 vulnerabilities apply to the implementation of variadic functions, where an argument count check can be trained to allow accessing out-of-bounds of the real arguments, or a stale argument can be accessed, both leading to pointer crafting.
\item \textbf{Interpreter dispatch loop.} The interpreter dispatch loop of a virtual machine typically consists of one or more indirect branches. The loop can be trained to speculatively transfer control to either the wrong bytecode (variant 1) or attacker-controlled code (variant 2), leading to pointer crafting.
\item \textbf{Indirect control transfers.} The variant 2 example can be exploited with indirect function calls in languages with first class functions or with virtual dispatch in object-oriented languages. Improper speculative control transfer leads to type confusion and pointer crafting.
\item \textbf{Call stack.} The CPU can be trained into mispredicting return addresses due to another internal predictor known as the \emph{return address stack}~\cite{SpectreReturns}, leading to pointer crafting.
\item \textbf{Switch statements.} Switch statements that are compiled to jump tables can be used to train the indirect branch predictor, either directly or through aliasing, to transfer control to code with improper type assumptions, leading to pointer crafting.
%REDACTED \item \textbf{Stack slots.} A typical implementation of the call stack includes space that is used to store arguments, temporary values, or registers that will be clobbered across calls. Due to variant 4, reads from the stack may speculatively produce stale values of the wrong type, leading to pointer crafting.
\end{enumerate}

In our offensive work for JavaScript and WebAssembly implementations, detailed in the next section, we developed proof-of-concept universal read gadgets using many of the above mechanisms.
As our examples show, our semantic model is able to capture the fundamental vulnerabilities that these gadgets rely upon, though architectural details determine exploitability.
Items near the top of the list generally require less knowledge than those at the bottom of the list. In particular, we found variant 1 to be quite simple to exploit. For managed languages, variant 3 is only different from variant 1 in that superuser memory can be accessed. Variant 2 is only easily exploitable if one has direct control over the virtual memory addresses of code. Variant 4 can be difficult to exploit reliably due to the black box nature of the memory disambiguator state. We focused exclusively on in-process attacks and not cross-process attacks.

\section{Mitigations}

\subsection{Disabling Speculation}

The most obvious mitigation against Spectre attacks is to explicitly disable speculation.
Unfortunately most CPUs don't provide such a mechanism, and even when they do, not to user-level programs.
On Intel CPUs the manufacturer recommendation is to use an \texttt{LFENCE} instruction to prevent speculation across this barrier as a protection for variant 1~\cite{IntelSideChannel}.

Unfortunately this approach has a number of limitations. First, the instruction was not designed as a speculation barrier, but as a memory load barrier. It prevents future speculative memory loads, and so mitigates side-channels that ex-filtrate data via the cache as all the examples in Section~\ref{sec:variants}. However the microarchitectural implementation details are proprietary and impossible to audit. It is not known what effects it really has. Indeed, this could be a problem if CPUs continue to speculatively execute other instructions, which still influence other \MuState and also consume execution time.
Second, this mitigation is local, and careful thought is required to insert \texttt{LFENCE} before every vulnerable operation. Variant 1 can be mitigated by inserting \texttt{LFENCE} before safety-check branches, but mitigating variant 4 may require insertion of an \texttt{LFENCE} before every vulnerable memory load operation, which is drastic. Because the mitigation is local, it cannot directly mitigate against variant 2, because an attacker with the power to speculatively jump anywhere can simply skip any inserted \texttt{LFENCE}. Third, this approach imposes an impractical performance overhead on execution. Requiring the CPU to stall its out-of-order execution pipeline before every branch (variant 1) or even every load operation (variant 4) reduces program performance by orders of magnitude as seen in Section~\ref{sec:production}.

\subsection{Timer Mitigation}

Individual optimizations based on $\mu$-state, such as a cache hit versus miss, give rise to extremely small differences in execution time which require a high resolution timer to detect. We might consider adjusting the precision of timers or removing them altogether as an attempt cripple the program's ability to read timing side-channels.

Unfortunately we now know that this mitigation is not comprehensive.
Previous research~\cite{FantasticTimers} has shown three problems with timer mitigations: (1) certain techniques to reduce timer resolution are vulnerable to resolution recovery, (2) timers are more pervasive than previously thought and (3) a high resolution timer can be constructed from concurrent shared memory. The Amplification Lemma from Section \ref{sec:amp-lemma} is the final nail in this coffin, as it shows (4) gadgets themselves can be amplified to increase the timing differences to arbitrary levels.

\subsection{Branchless Masking} \label{sec:branchless}

As we've now seen in detail, the variant 1 vulnerability means that any safety check inserted as a branch into the code has the potential to be bypassed by the inherent nature of branch prediction.
This is captured in our semantic model in that the reorder buffer does not encode control dependencies between a branch and instructions following that branch, allowing them to execute out of order with respect to each other.
Our model therefore informed the design of our mitigations, since it forced us to assume that branches can no longer be trusted as security mechanisms in speculation. This realization led us to designing speculative safety checks that do not rely on branches.

\subsubsection{Array index masking} \label{sec:array-index-masking}

The intuition behind \emph{array index masking} is to automatically insert additional arithmetic computations to force attacker-supplied out-of-bounds indices to be within bounds, even in misspeculated execution. In the case of array accesses, the safety check already computes the expression \texttt{index >= array.length}. In our model, this condition is computed via a \IBinop \Binop{ge} instruction that produces either \Zero or \One in a register, which is then used as input to a \IBranch instruction. Following the branch instruction, on the success (in-bounds) path, we simply insert a multiplication of the index by the inverse of the condition, so that if the condition is true (i.e., the index is out-of-bounds), the attacker-chosen index is clamped to \Zero. Figure~\ref{fig:mitigate-var1} shows the machine code that would be generated for Figure~\ref{subfig:var1code} with this mitigation enabled.

\begin{figure}
\tiny {
\centering
\begin{minipage}{.5\textwidth}
  \centering
  \begin{lstlisting}
@mitigated
 load r1 [#array - 1]   ; load array length
 binop[ge] r2 r0 r1     ; index >= array length
 branch r2 @oob         ; (vuln) bounds check
 const r3 #1            ; (mitig) 
 binop[xor] r2 r2 r3    ; (mitig) invert condition
 binop[mul] r0 r0 r2    ; (mitig) index := 0 if oob
 load r0 [r0 + #array]  ; load [array + index]
 load r0 [r0 + #timing] ; load [timing + val]
 jump rip               ; return
@oob:
 const r0 #0            ; out of bounds => 0
 jump rip               ; return
  \end{lstlisting}
  \captionof{figure}{Branchless array index masking mitigation}
  \label{fig:mitigate-var1}
\end{minipage}%
\begin{minipage}{.5\textwidth}
  \centering
  \begin{lstlisting}
@program_start:
 const rp 1           ; Initialize poison register
    ...
 branch rc @target    ; (vuln) branch
 const rt 1           ; (mitig) 
 binop[xor] rt rt rc  ; (mitig) rt is inverted rc
 binop[and] rp rp rt  ; (mitig) set rp to 0 if rc == 1
    ...
@target:
 binop[and] rp rp rc  ; (mitig) set r0 to 0 if rc == 0
    ...
 binop[mul] rt rp r0  ; (mitig) mask load addr with rp
 load r0 [rt]         ; load array
  \end{lstlisting}
  \captionof{figure}{Pervasive conditional masking}
  \label{fig:mitigate-var1-pervasive}
\end{minipage}
} % tiny
\end{figure}

The extra calculation is always a no-op in the architected path, so it strictly adds overhead, not only in code size, but in execution time. There is also a risk to being a no-op on the architected path, as an aggressively optimizing \MuArch might even compile it away, reintroducing the original vulnerability in the pursuit of performance. In fact, in the course of our practical mitigation work in V8, an engineer introduced an optimization in the backend of V8's optimizing compiler that unwittingly removed Spectre mitigations inserted by the frontend.

%TODO put in array masking expression
%TODO performance numbers
%TODO mention WebAssembly memory masking

%We can now prove that this mitigation prevents accessing arbitrary out-of-bounds memory. We use our model of speculation and a simple brute-force enumeration of all the potential \MuStates for the program. There are only 3 cases.

\subsubsection{Pervasive conditional masking}

Array index accesses are just one of many types of safety checks inserted by language implementations. As we've seen, type checks, argument count checks, and others can constitute potential vulnerabilities. The logical extension of the array index masking technique is to use the conditions computed for any of these branches to compute a \emph{poison} value that is used as an additional input to all loads that are control-dependent on the safety-check.

To implement general conditional masking, we reserve one otherwise unused register, denoted by $r_p$, for the poison value.
The register is initially set to $1$, a value it will have whenever the processor is correctly speculating.
We will instrument every branch target to update the poison register so that a misspeculation can still execute, but the poison register will be set to $0$, so that the poison value can be used to destroy any sensitive information in misspeculation.
In effect, we compute a redundant condition in dataflow in addition to control flow.
In the true branch, we will mask the poison register with the condition register so that the poison register becomes $0$ if the condition was $0$ and in the false branch we set the poison register is set to $0$ if the condition is $1$.
After instrumenting the branches to add to the running speculative poison, we will instrument all loads to consume the poison by masking every load's address with the poison register.  An example of the generated code is shown in Figure~\ref{fig:mitigate-var1-pervasive}.

\textbf{Proof sketch.} To show correctness of this instrumentation, assuming that loads from address $0$ are not secret, we will argue that if an execution in the speculative semantics (Section~\ref{sec:modeling-control-speculation}) loads memory location $l$, then there is an execution in the architecture that reads the same location. If the speculative execution reaches a load of memory location $l$ and the load occurs in the reorder buffer before any misspeculation, then Theorem \ref{thm:spec-arch-agree} guarantees the load is also reachable in the architecture model. If the load occurs after a mis-speculated branch in the reorder buffer, the instructions sequence following the branch must have updated the poison register to $0$ and the load's address must have been multiplied with the poison, thus loading from address $0$.

\subsubsection{Pervasive indirect branch masking}

To guard type safety across indirect branches, such as bytecode handler dispatch and virtual function dispatch, we need to extend masking even further. A poison is computed by comparing the expected target of the indirect branch with the reached target. To do so, we reserve a register $r_b$ to used for the target address of all indirect branches. All potential targets of indirect branches are instrumented to perform a branchless comparison of the current \texttt{pc} with $r_b$, and use the result of this comparison to update the poison register $r_p$. 

Note that this mitigation strategy does not prevent mispredicted indirect branch speculation, but instead clears the poison register in the case of misspeculation, and relies on the masking of loads to prevent the speculative execution from accessing memory that could expose secret data. It also only prevents variant 1 indirect branch attacks. If an attacker can exploit a CPUs approximate prediction state, as in a variant 2 attack, they can train an indirect branch to alias with a location that isn't a indirect branch target, and therefore isn't instrumented with the indirect target poision computation and doesn't reset the poison register.

\subsection{Mitigating Other Mispredictions}

All previous mitigations rely on specially-crafted code to avoid either accessing or leaking information in misspeculation.
Variant 2 unfortunately shows that some predictions allow attackers to take paths that aren't architecturally possible.
Those paths are not under our control and hence cannot be made to include mitigations.
The only solution is to avoid such predictors altogether.
This has led to the suggested use of \emph{retpoline}~\cite{IntelRetpoline}, a construct that avoids using indirect branches altogether to block predictions from the \emph{branch target buffer}.
Unfortunately, this technique actually makes use of a \emph{second} predictor, the return stack buffer, which happens to take priority on Intel CPUs.
Retpolines do not work on other CPU models or architectures.

Variant 4 shows that loads can sometimes observe completely wrong values in speculation. This can lead to pointer crafting or using wrong addresses for indirect jumps. There is no real known way to mitigate this in software. Hardware manufacturers have provided mitigations in the form of a new off-by-default mitigation by Intel~\cite{IntelSideChannel} and AMD~\cite{AMDSpeculative}, and two new barriers SSBB and PSSBB from ARM~\cite{ARMSideChannel} that block loads from bypassing stores to the same virtual and physical address respectively.

\subsection{Implementation in a Production JavaScript VM} \label{sec:production}

As part of our offensive work, we developed proofs of concept in C++, JavaScript, and WebAssembly for all the reported vulnerabilities.
We were able to leak over 1KB/s from variant 1 gadgets in C++ using \Binop{rdtsc} with 99.99\% accuracy and over 10B/s from JavaScript using a low resolution timer.
We demonstrated a potential 2.5KB/s variant 4 vunerability, but with low reliability, starting at 0.01\% but amplifiable up to 20\% through various techniques.
We found that using shared memory to construct a timer worked well enough in JavaScript to measure individual cache hits and misses and exploit any of the known leaks.

As part of our defensive work, we implemented a number of the described mitigations in the V8 JavaScript virtual machine and evaluated their performance penalties. As we've noted, none of these mitigations provide comprehensive protection against Spectre, and so the mitigation space is a frustrating performance / protection trade-off.

We augmented every branch with an \texttt{LFENCE} instruction, which provides variant 1 protection\footnote{Although not variant 4, which would require an \texttt{LFENCE} before every memory load}. The overhead was considerable, with a 2.8x slowdown on the Octane JavaScript benchmark (some line items up to 5x). We also experimented with the \emph{retpoline} code sequence, inserting it into every generated indirect jump to protect against variant 2. Due to the dynamic nature of JavaScript, V8 emits a significant number of indirect jumps and so \emph{retpoline} also has a high performance penalty of 1.52x slowdown on Octane. Disabling the emission of jump tables for WebAssembly has highly variable costs of up to 5x, but retpoline for all indirect calls costs only 0-2\%.

In order to provide more performant mitigations, we took a tactical approach to target specific vulnerabilities. Implementing array masking incurs a 10\% slowdown on Octane, while the more pervasive conditional poisoning using a reserved poison register (which protects against variant 1 type confusion) incurs a 19\% slowdown on Octane. In addition we implemented pervasive indirect branch masking on the V8 interpreter's dynamic dispatch and on indirect function calls. This incurred negligible overhead on code optimized by V8's optimizing compiler, however with the optimizer disabled it incurs a 12\% slowdown on V8's interpreter when running Octane. For WebAssembly, we implemented unconditional memory masking (by padding the memory to a power-of-2 size and always loading a mask), which incurs a 10-20\% slowdown. For variant 4, we implemented a mitigation to zero the unused memory of the heap prior to allocation, which cost about 1\% when done concurrently and 4\% for scavenging.

Variant 4 defeats everything we could think of.
We explored more mitigations for variant 4 but the threat proved to be more pervasive and dangerous than we anticipated.
For example, stack slots used by the register allocator in the optimizing compiler could be subject to type confusion, leading to pointer crafting.
Mitigating type confusion for stack slots alone would have required a complete redesign of the backend of the optimizing compiler, perhaps man years of work, without a guarantee of completeness.
We recognized quickly that a compiler backend overhaul, a complete audit of the entire runtime system, and application of (not yet designed) mitigations in the C++ compiler for the VM's code itself were intractable for essentially any-sized codebase.
For this reason we do not believe that variant 4 can be effectively mitigated in software, due not just to manpower, but a lack of architectural options, since reasoning about variant 4 requires the confounding assumption that \emph{in speculation, writes to memory may not be visible to subsequent reads at all}.

A subset of the implemented mitigations shipped in successive Chrome releases from 64 to 67 and formed the initial primary defensive strategy. Timer mitigations were also shipped early, since they were an easy way to slow attackers. Accordingly, \texttt{SharedArrayBuffer} which represents concurrent shared memory and can be used to construct a timer, was disabled. In recognition of the fact that software mitigation is still an open problem for virtual machines, and that mitigations would also need to be applied to all of the millions of lines of C++ code in the browser, Chrome's defensive strategy shifted entirely to site isolation~\cite{SiteIsolation}, which sandboxes code from different origins in different processes, thus relying on hardware-enforced protection.

% \section{What we're missing}
% Mitigation for variant 1 only covers side-effects to cache \MuState, not branch prediction, aliasing, or timing itself.
% Mitigation doesn't cover information stored in branch prediction itself (e.g. an attacker trying to harvest information from the branch predictors).
% RSB vulnerabilities waiting.
% Out-of-bound writes causing broken metadata (even if the memory was read-only).
% Other \MuState waiting to be discovered: cache bandwidth, frequency scaling, functional unit occupancy, cache coherency state, TLB, MMU, store buffer
% Not known if there are more cross-process leaks.
% Value speculation would defeat software mitigations and open a new class of vulnerabilities.
% Some hardware mitigations are probably necessary.
% Store forwarding is already a type of value speculation (on the address stored to/loaded from).
% Proofs are only as good as models they are based on.

\section{Conclusion} \label{sec:conclusion}

Spectre defeats an important layer of software security.
The community has assumed for decades that programming language security enforced with static and dynamic checks could guarantee confidentiality between computations in the same address space.
Our work has discovered there are numerous vulnerabilities in today's languages that when run on today's CPUs allow construction of the universal read gadget, which completely destroys language-enforced confidentiality.
This paper has attempted to shed light on side-channels that have been hiding in simulators and CPUs, a fact that has received little attention in the past and is only now coming to the fore.
Hardware/OS process isolation is needed now more than ever, as we predict that language mechanisms to enforce confidentiality will be constantly threatened by side-channels.
From our first attempts at modeling recently-disclosed vulnerabilities to our work on software mitigations, it has become painfully obvious to us that we are facing three massive open problems:
\begin{enumerate}
\item Finding \MuArchal side channels requires enumerating and modeling relevant \MuState, a difficult task for processors that are closed source and full of valuable and carefully-guarded intellectual property.
\item Understanding vulnerabilities requires us to model how programs can manipulate and observe \MuState, which also requires us to understand complex \MuState in black-box processors.
\item Mitigating vulnerabilities is perhaps the most challenging of all, since efficient software mitigations needed for extant hardware seem to be in their infancy, and hardware mitigation for future designs is a completely open design problem.
\end{enumerate}

Computer systems have become massively complex in pursuit of the seemingly number-one goal of performance.
We've been extraordinarily successful at making them faster and more powerful, but also more complicated, facilitated by our many ways of creating abstractions.
The tower of abstractions has allowed us to gain confidence in our designs through separate reasoning and verification, separating hardware from software, and introducing security boundaries.
But we see again that our abstractions leak, side-channels exist \emph{outside} of our models, and now, down deep in the hardware where we were not supposed to see, there are vulnerabilities in the very chips we deployed the world over.
Our models, our \emph{mental} models, are wrong; we have been trading security for performance and complexity all along and didn't know it.
It is now a painful irony that today, defense requires even more complexity with software mitigations, most of which we know to be incomplete.
And complexity makes these three open problems all that much harder.
Spectre is perhaps, too appropriately named, as it seems destined to haunt us for a long time.

%% Acknowledgments
\section*{Acknowledgements}
Our work on Spectre was a close collaboration among dozens of engineers across several companies. We would especially like to acknowledge  Jann Horn, Matt Linton, Chandler Carruth, Paul Turner, Michael Starzinger, Brad Nelson, Chris Palmer, Justin Schuh, and Charlie Reis at Google, Luke Wagner at Mozilla, Filip Pizlo, Michael Saboff and Robin Morrisset at Apple, John Hazen and Louis Lafreniere at Microsoft, Jason Brandt at Intel, Alastair Reid and Rodolph Perfetta at ARM.

%% Bibliography
\bibliography{bib}
\bibliographystyle{plain}

%% Appendix
\appendix
\section{Appendix}

\subsection{Variant 2: Illustrative Example}\label{app:var2}

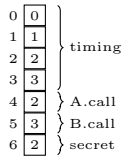
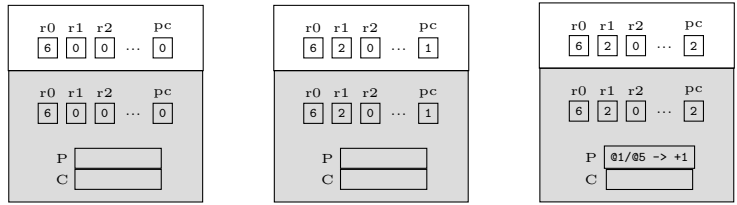
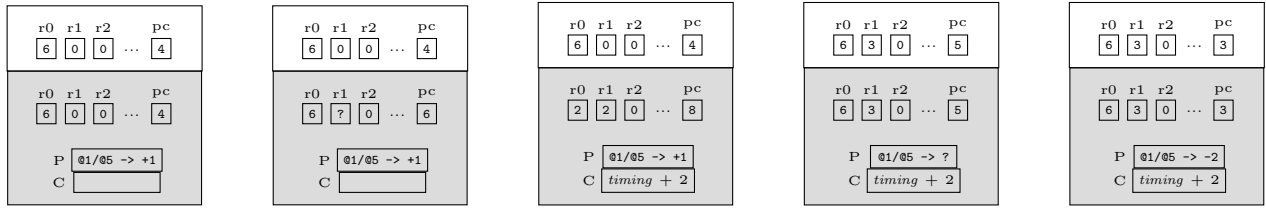
\begin{figure}[p]
{\tiny
\centering
\begin{subfigure}{0.4\textwidth} 
  \centering
  \begin{lstlisting}
function train(a, arg) {
  return a.call(arg);
}
function A.call(arg) { return arg; }
function B.call(arg) { return arg; }
function vulnerable(b, arg) {
  return b.call(arg);
}
function attack(arg) {
  return timing[*arg];
}

int addr = <craft secret address>;
train(new A(), addr);   // Train
attack(new B(), addr);  // Attack
  \end{lstlisting}
  \caption{Vulnerable code}
\end{subfigure}
~
\begin{subfigure}{0.5\textwidth}  
  \centering
  \begin{lstlisting}
@train:
 0: load r1 [r0 + #0]       ; load a.call
 1: jump r1                 ; (vulnerable) indirect jump
@A.call:
 2: jump r15                ; return
@B.call:
 3: jump r15                ; return

@vulnerable:
 4: load r1 [r0 + #0]       ; load a.call
 5: jump r1                 ; (vulnerable) indirect jump
@attack:
 6: load r1 [r0]            ; load *arg
 7: load r0 [r1 + #timing]  ; load [timing + val]
 8: jump r15                ; return
  \end{lstlisting}
  \caption{Vulnerable machine code}
\end{subfigure}

\hspace{-15ex}
\begin{subfigure}{0.3\textwidth}
  \centering
  \vspace{-20ex}
  \begin{tikzpicture}[]
  \matrix(memory)[column sep=0mm, ampersand replacement=\&] {
    \node(){0}; \& \node(memTimingStart)[draw]{0}; \\ 
    \node(){1}; \& \node()[draw]{1}; \\ 
    \node(){2}; \& \node()[draw]{2}; \\
    \node(){3}; \& \node(memTimingEnd)[draw]{3}; \\ 
    \node(){4}; \& \node(AVal)[draw]{2}; \\
    \node(){5}; \& \node(BVal)[draw]{3}; \\
    \node(){6}; \& \node(secret)[draw]{2}; \\
  };
  \memLabel{timing}{memTimingStart}{memTimingEnd}
  \memLabel{A.call}{AVal}{AVal}
  \memLabel{B.call}{BVal}{BVal}
  \memLabel{secret}{secret}{secret}
  \end{tikzpicture}
\caption{Memory layout}
\end{subfigure}
~
\begin{subfigure}[t]{0.6\textwidth}
\mbox{
  \realArchState{6}{0}{0}{0}
  \uArchState{6}{0}{0}{0}
  \cacheState{}{}
  \drawCpu
~
  \realArchState{6}{2}{0}{1}
  \uArchState{6}{2}{0}{1}
  \cacheState{}{}
  \drawCpu
~
  \realArchState{6}{2}{0}{2}
  \uArchState{6}{2}{0}{2}
  \cacheState{\texttt{@1/@5 -> +1}}{}
  \drawCpu
}
\vspace{-0.75ex}
\caption{Training the predictor}
\label{subfig:var2train}
\end{subfigure}

\hspace{-15ex}
\begin{subfigure}[t]{1\textwidth}
\mbox{
  \realArchState{6}{0}{0}{4}
  \uArchState{6}{0}{0}{4}
  \cacheState{\texttt{@1/@5 -> +1}}{}
  \drawCpu
~
  \realArchState{6}{0}{0}{4}
  \uArchState{6}{?}{0}{6}
  \cacheState{\texttt{@1/@5 -> +1}}{}
  \drawCpu
~
  \realArchState{6}{0}{0}{4}
  \uArchState{2}{2}{0}{8}
  \cacheState{\texttt{@1/@5 -> +1}}{\emph{timing} $+$ 2}
  \drawCpu
~
  \realArchState{6}{3}{0}{5}
  \uArchState{6}{3}{0}{5}
  \cacheState{\texttt{@1/@5 -> ?}}{\emph{timing} $+$ 2}
  \drawCpu
~
  \realArchState{6}{3}{0}{3}
  \uArchState{6}{3}{0}{3}
  \cacheState{\texttt{@1/@5 -> -2}}{\emph{timing} $+$ 2}
  \drawCpu
}
\caption{Performing the attack}
\label{subfig:var2attack}
\end{subfigure}
}  % /tiny
\caption{Illustration of a variant 2 attack}
\label{fig:var2}
\end{figure}

The example shown in Figure~\ref{fig:var2} illustrates the variant 2 vulnerability where target misreconstruction is used to leak secret data. The \texttt{train} routine calls a virtual function via an indirect jump to train the branch predictor. However the branch predictor only stores the lower 2 bits of the source address and encodes the target as an offset (Figure~\ref{subfig:var2train}). As such, the prediction is ambiguous and predicts an indirect jump from \texttt{@5->@6}, as well as the intended \texttt{@1->@2}. An attacker takes advantage of this ambiguity by positioning the \texttt{attack} routine on the target side of this misprediction, and calling \texttt{vulnerable}. The CPU speculative executes the \texttt{attack} routine instead of the intended \texttt{B.call}, which as before speculatively loads an \emph{attacker-crafted pointer} and encodes that value into the cache

\subsection{Variant 4: Illustrative Example}\label{app:var4}

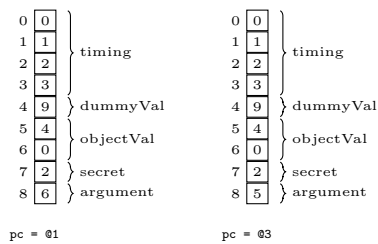
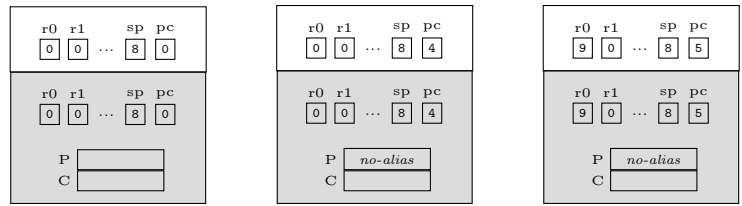
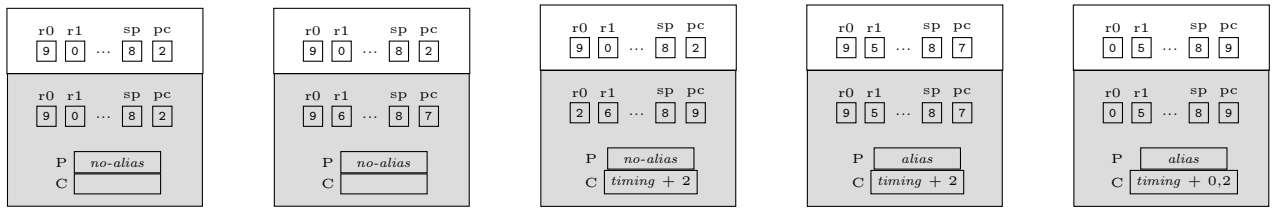
\begin{figure}[p]
{\tiny
\centering
\begin{subfigure}{0.45\textwidth} 
  \centering
  \begin{lstlisting}
function dummyInt(argument) {
  return *dummyVal;
}
function loadObject(argument) {
  var obj = *argument;
  return timing[obj.f]
}

function vulnerable() {
  // Train
  var argument = <craft secret address - 1>;
  loadInt(&argument);
  // Attack
  argument = objectN;
  loadObject(&argument);
}
  \end{lstlisting}
  \caption{Vulnerable code}
\end{subfigure}
~
\begin{subfigure}{0.5\textwidth}  
  \centering
  \begin{lstlisting}
@vulnerable:
 0: store [sp] #(secret_addr-1)  ; pass argument on stack
 1: call @dummyInt               ; call dummyInt()
 2: store [sp] #objectVal        ; (vulnerable) update stack
 3: call @loadObject             ; call loadObject()

@dummyInt
 4: load r0 [#dummyVal]          ; load dummyVal
 5: jump r15                     ; return

@loadObject:  
 6: load r1 [sp + #0]            ; load argument
 7: load r0 [r1 + #1]            ; (vulnerable) load obj.f
 8: load r0 [r0 + #timing]       ; timing disclosure
 9: jump r15                     ; return
  \end{lstlisting}
  \caption{Vulnerable machine code}
\end{subfigure}

\hspace{-15ex}
\begin{subfigure}{0.4\textwidth}
  \centering
  \vspace{-20ex}
  \begin{tikzpicture}[]
  \matrix(memory)[column sep=0mm, ampersand replacement=\&] {
    \node(){0}; \& \node(memTimingStart)[draw]{0}; \\ 
    \node(){1}; \& \node()[draw]{1}; \\ 
    \node(){2}; \& \node()[draw]{2}; \\
    \node(){3}; \& \node(memTimingEnd)[draw]{3}; \\ 
    \node(){4}; \& \node(memDummy)[draw]{9}; \\
    \node(){5}; \& \node(memObjStart)[draw]{4}; \\
    \node(){6}; \& \node(memObjEnd)[draw]{0}; \\
    \node(){7}; \& \node(memSecret)[draw]{2}; \\
    \node(){8}; \& \node(memStack)[draw]{6}; \\
  };
  \memLabel{timing}{memTimingStart}{memTimingEnd}
  \memLabel{objectVal}{memObjStart}{memObjEnd}
  \memLabel{dummyVal}{memDummy}{memDummy}
  \memLabel{secret}{memSecret}{memSecret}
  \memLabel{argument}{memStack}{memStack}
  \node()[fit=(memory.south east)(memory.south west), yshift=-5ex,
          minimum width=20ex, font=\ttfamily]{pc~=~@1};
  \end{tikzpicture}
  ~
  \begin{tikzpicture}[]
  \matrix(memory)[column sep=0mm, ampersand replacement=\&] {
    \node(){0}; \& \node(memTimingStart)[draw]{0}; \\ 
    \node(){1}; \& \node()[draw]{1}; \\ 
    \node(){2}; \& \node()[draw]{2}; \\
    \node(){3}; \& \node(memTimingEnd)[draw]{3}; \\ 
    \node(){4}; \& \node(memDummy)[draw]{9}; \\
    \node(){5}; \& \node(memObjStart)[draw]{4}; \\
    \node(){6}; \& \node(memObjEnd)[draw]{0}; \\
    \node(){7}; \& \node(memSecret)[draw]{2}; \\
    \node(){8}; \& \node(memStack)[draw]{5}; \\
  };
  \memLabel{timing}{memTimingStart}{memTimingEnd}
  \memLabel{objectVal}{memObjStart}{memObjEnd}
  \memLabel{dummyVal}{memDummy}{memDummy}
  \memLabel{secret}{memSecret}{memSecret}
  \memLabel{argument}{memStack}{memStack}
  \node()[fit=(memory.south east)(memory.south west), yshift=-5ex,
          minimum width=20ex, font=\ttfamily]{pc~=~@3};
  \end{tikzpicture}
\caption{Memory layout}
\end{subfigure}
~
\begin{subfigure}[t]{0.6\textwidth}
\mbox{
  \realArchStateSp{0}{0}{8}{0}
  \uArchStateSp{0}{0}{8}{0}
  \cacheState{}{}
  \drawCpu
~
  \realArchStateSp{0}{0}{8}{4}
  \uArchStateSp{0}{0}{8}{4}
  \cacheState{\emph{no-alias}}{}
  \drawCpu
~
  \realArchStateSp{9}{0}{8}{5}
  \uArchStateSp{9}{0}{8}{5}
  \cacheState{\emph{no-alias}}{}
  \drawCpu
}
\caption{Training the predictor}
\label{subfig:var4train}
\end{subfigure}

\hspace{-15ex}
\begin{subfigure}[t]{1\textwidth}
\mbox{
  \realArchStateSp{9}{0}{8}{2}
  \uArchStateSp{9}{0}{8}{2}
  \cacheState{\emph{no-alias}}{}
  \drawCpu
~
  \realArchStateSp{9}{0}{8}{2}
  \uArchStateSp{9}{6}{8}{7}
  \cacheState{\emph{no-alias}}{}
  \drawCpu
~
  \realArchStateSp{9}{0}{8}{2}
  \uArchStateSp{2}{6}{8}{9}
  \cacheState{\emph{no-alias}}{\emph{timing} $+$ 2}
  \drawCpu
~
  \realArchStateSp{9}{5}{8}{7}
  \uArchStateSp{9}{5}{8}{7}
  \cacheState{\emph{alias}}{\emph{timing} $+$ 2}
  \drawCpu
~
  \realArchStateSp{0}{5}{8}{9}
  \uArchStateSp{0}{5}{8}{9}
  \cacheState{\emph{alias}}{\emph{timing} $+$ 0,2}
  \drawCpu
}
\caption{Performing the attack}
\label{subfig:var4attack}
\end{subfigure}
}  % /tiny
\caption{Illustration of a variant 4 attack}
\label{fig:var4}
\end{figure}

The example in Figure~\ref{fig:var4} represents an example of a variant 4, speculative aliasing confusion attack.  A memory location (e.g., a stack slot) is reused to pass \texttt{argument} to two different functions which each expect \texttt{argument} to be of a different type. The attacker first repeatedly stores an integer into \texttt{argument} and calls \texttt{dummyInt}. Since this routine doesn't load the value in \texttt{argument}, the memory disambiguator will predict that stores to \texttt{argument} have \emph{no-alias} with subsequent stores (Figure~\ref{subfig:var4train}). The attacker then overwrites \texttt{argument} with a pointer to an object and calls \texttt{loadObject}. The memory disambiguator mispredicts that subsequent loads depend on this store, and therefore speculatively executes \texttt{loadObject} before the store has completed (Figure~\ref{subfig:var4attack}). As a result, it incorrectly treats an attacker-controlled integer as an object pointer, which, as before, can be used to load \x{secret} memory state that can be exfiltrated from speculation via a timing array, encoding it into the \MuState of the cache.

\end{document}